\documentclass{article}

\usepackage[square,numbers]{natbib}
\usepackage[preprint]{neurips_2025}

\usepackage[utf8]{inputenc} 
\usepackage[T1]{fontenc}    
\usepackage{hyperref}       
\usepackage{url}            
\usepackage{booktabs}       
\usepackage{amsfonts}       
\usepackage{nicefrac}       
\usepackage{microtype}      
\usepackage{xcolor}         
\usepackage{graphicx}
\usepackage{amsmath}
\usepackage{float} 
\usepackage{amssymb}

\usepackage{adjustbox}
\usepackage{graphicx} 
\usepackage{pifont}  
\usepackage{enumitem}

\newtheorem{definition}{Definition}

\title{Scale-Invariance Drives Convergence in AI and Brain Representations}

\author{%
Junjie Yu$^{1}$ \thanks{J. Yu and W. Ma contributed equally.} \quad Wenxiao Ma$^{1\ *}$ \quad Jianyu Zhang$^1$ \quad Haotian Deng$^1$ \quad Zihan Deng$^1$ \\ \textbf{Yi Guo}$^{2\ \dag}$  \quad \textbf{Quanying Liu}$^1$ \thanks{Corresponding author.} \\
$^1$Department of Biomedical Engineering, Southern University of Science and Technology \\ $^2$Shenzhen People's Hospital, affiliated to Southern University of Science and Technology\\
\texttt{\{12231192, 12212659\}@mail.sustech.edu.cn}\\
\texttt{xuanyi\_guo@163.com}\\
\texttt{liuqy@sustech.edu.cn}
}

\begin{document}

\maketitle


\begin{abstract}
Despite variations in architecture and pretraining strategies, recent studies indicate that large-scale AI models often converge toward similar internal representations that also align with neural activity. We propose that scale-invariance, a fundamental structural principle in natural systems, is a key driver of this convergence. In this work, we propose a multi-scale analytical framework to quantify two core aspects of scale-invariance in AI representations: dimensional stability and structural similarity across scales. We further investigate whether these properties can predict alignment performance with functional Magnetic Resonance Imaging (fMRI) responses in the visual cortex. Our analysis reveals that embeddings with more consistent dimension and higher structural similarity across scales align better with fMRI data. Furthermore, we find that the manifold structure of fMRI data is more concentrated, with most features dissipating at smaller scales. Embeddings with similar scale patterns align more closely with fMRI data. We also show that larger pretraining datasets and the inclusion of language modalities enhance the scale-invariance properties of embeddings, further improving neural alignment. Our findings indicate that scale-invariance is a fundamental structural principle that bridges artificial and biological representations, providing a new framework for evaluating the structural quality of human-like AI systems.
\end{abstract}

\section{Introduction}

Large-scale AI models, despite differences in architecture and training strategies, often converge to similar internal representational structures \citep{huh2024platonic}. This phenomenon is observed not only across models within the same modality but also across different modalities and even distinct systems. Prior studies have demonstrated that model embeddings, even without explicit alignment, can effectively predict brain activity \citep{wang2023better, contier2024distributed, shen2024towards, demircan2024evaluating}.

We hypothesize that \textbf{this convergence reflects deeper structural regularities inherent in the external world}. As models absorb more data, these regularities shape internal representations, leading to shared geometric organization across models. Identifying these structural regularities can offer valuable insights into the principles underlying both artificial and biological systems.

\begin{figure}[H]
  \centering
  \includegraphics[width=0.98\linewidth]{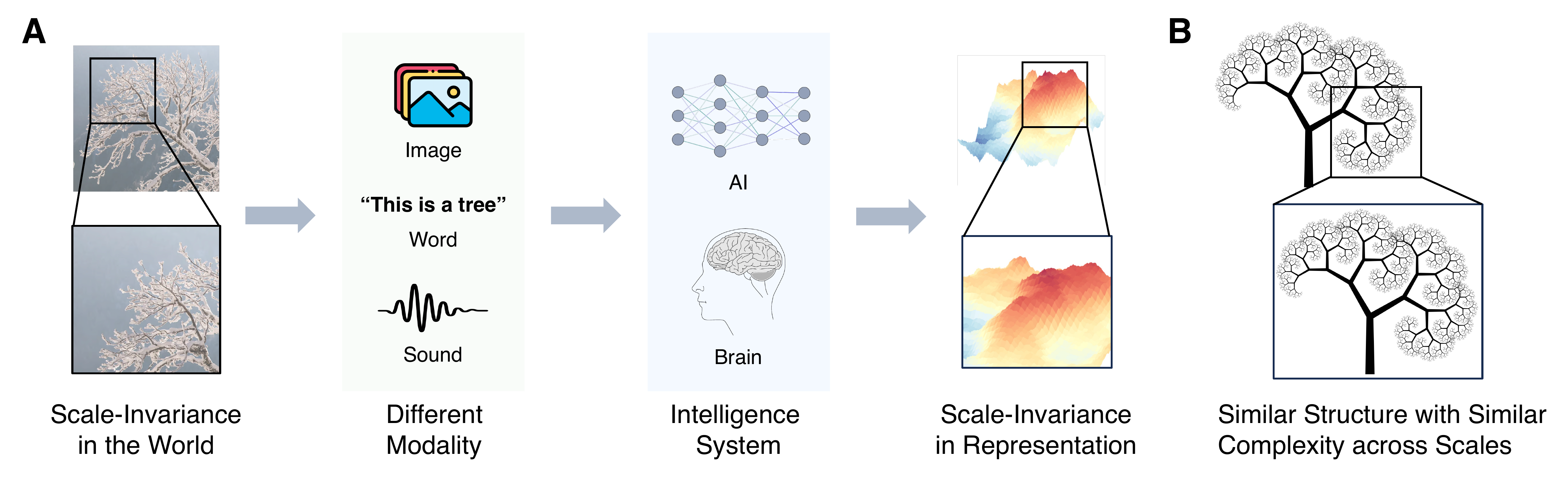}
  \caption{\textbf{Scale-Invariance as a Structural Bridge Between AI and Neural Representations}. \textbf{A}. External structural constraints, such as scale-invariance, shape the formation of representational manifolds in both AI and brain systems. \textbf{B}. Structural scale-invariance is defined by self-similarity and consistent complexity across scales.}
  \label{fig:idea}
\end{figure}

To investigate these regularities, we focus on the concept of scale-invariance, a phenomenon widely observed across various domains, including physical, biological and cognitive systems \citep{stanley2000scale, nakayama2015scale, dymarsky2016scale}. Scale-invariance refers to the stability of properties across different scales, as seen in fractal structures, power-law distributions in language \citep{mehri2016power} and neural activity \citep{grosu2023fractal}. We hypothesize that \textbf{as AI models absorb more data, external structural constraints become more pronounced in their internal representations, aligning them more closely with neural representations} (Figure \ref{fig:idea} A). To validate this hypothesis, we investigate whether the degree of scale-invariance in structure of embedding manifold can effectively predict the extent of alignment with neural activity.

In this context, we define structural scale-invariance as comprising two core properties (Figure \ref{fig:idea} B):

\begin{itemize}[leftmargin=*]
    \item \textbf{Dimensional Stability}: Quantifies the consistency of structural complexity across scales.
    \item \textbf{Structural Self-Similarity}: Measures the similarity of geometric structures across scales.
\end{itemize}

To assess the predictive power of these properties, we examine whether dimensional stability and structural self-similarity in embeddings can effectively predict alignment with fMRI data. This analysis provides a comprehensive framework for characterizing structural organization across predefined scales.

However, due to constraints related to data acquisition and resolution, our analysis of scale-invariance is confined to finite scales. This limitation restricts our ability to capture structural patterns that may persist at very small scales. To address this gap, we extend our analysis using persistent homology \citep{edelsbrunner2008persistent, zomorodian2004computing}, which quantifies the distribution of structural features across a broader range of scales. This approach allows us to assess whether structural features are concentrated within specific scales or distributed more broadly, offering a more comprehensive perspective on neural manifold organization.

Our findings reveal that structural features in neural manifolds are not uniformly distributed across infinitesimal scales but are instead concentrated within a compact range. This compact distribution suggests that neural manifolds maintain structural efficiency, reducing the need for extensive analysis at very small scales.

Based on these analysis, we report three key findings:

\begin{itemize}[leftmargin=*]
    \item \textbf{Structural Consistency Across Scales}: Embeddings with higher dimensional stability and stronger self-similarity across scales exhibit more effective alignment with fMRI data, underscoring the importance of preserving structural consistency in capturing brain-like representations.
    \item \textbf{Compact Structural Distribution}: Neural manifolds display compact scale distributions, with structural features concentrated within specific scale ranges rather than extending to infinitesimal scales. Embeddings that replicate this compact distribution achieve stronger alignment with neural data, highlighting compactness as a critical aspect of structural efficiency.
    \item \textbf{Enhanced Scale-Invariance via Pre-training and Modality Integration}: Incorporating language modalities and pre-training on larger datasets amplifies the scale-invariance properties of embeddings, further improving their alignment with neural structures.
\end{itemize}

Collectively, these findings position scale-invariance as a unifying structural principle that not only drives the convergence of AI and neural representations but also offers a systematic framework for evaluating the multi-scale organization of high-dimensional embeddings.

\section{Related Work}

\paragraph{Alignment between AI and Brain}
Large-scale pretrained models have shown strong alignment with neural activity, particularly when trained on diverse, extensive datasets \citep{wang2023better, contier2024distributed}. Beyond evaluation, alignment also reveals insights into brain function \citep{waldrop2024can}. For instance, the alignment of autoregressive language models with fMRI data suggests that the brain may employ similar predictive mechanisms to model language \citep{schrimpf2021neural}.

\paragraph{Scale-Invariance in Natural and Neural Systems.}
Scale-invariance reflects self-similarity across scales. In nature, it manifests in fractal structures such as mountain ranges, tree branching and coastlines \citep{brown2002fractal, mandelbrot1998nature, tarboton1988fractal}. In neural systems, scale-invariance is observed in the fractal-like patterns of cortical folding, the branching structures of neuronal dendrites, and the power-law distribution of neural connectivity \citep{grosu2023fractal, wang2019human, liao2023topology}. On an abstract level, it appears in structure of human language \citep{mehri2016power} and in social networks \citep{zang2018power}. Despite its ubiquity, the role of scale-invariance in AI embeddings and its implications for neural alignment remain underexplored.

\paragraph{Methods for Analyzing Scale-Invariance.}
Scale-invariance is typically assessed using power-law fitting and fractal dimension analysis \citep{clauset2009power, barabasi1999emergence, proekt2012scale}, which capture self-similarity but do not quantitatively assess its extent. This limitation underscores the need for more robust frameworks capable of systematically quantifying scale-invariance across multiple scales

\section{Preliminaries and Technical Background}

Our methodological framework follows three key steps: (1) aligning pretrained model embeddings with fMRI responses under visual stimuli, (2) assessing the predictive power of scale-invariance properties of embedding for alignment, and (3) quantitatively characterizing the distribution of topological structures using topological analysis.

\begin{figure}[htbp]
  \centering
  \includegraphics[width=0.99\linewidth]{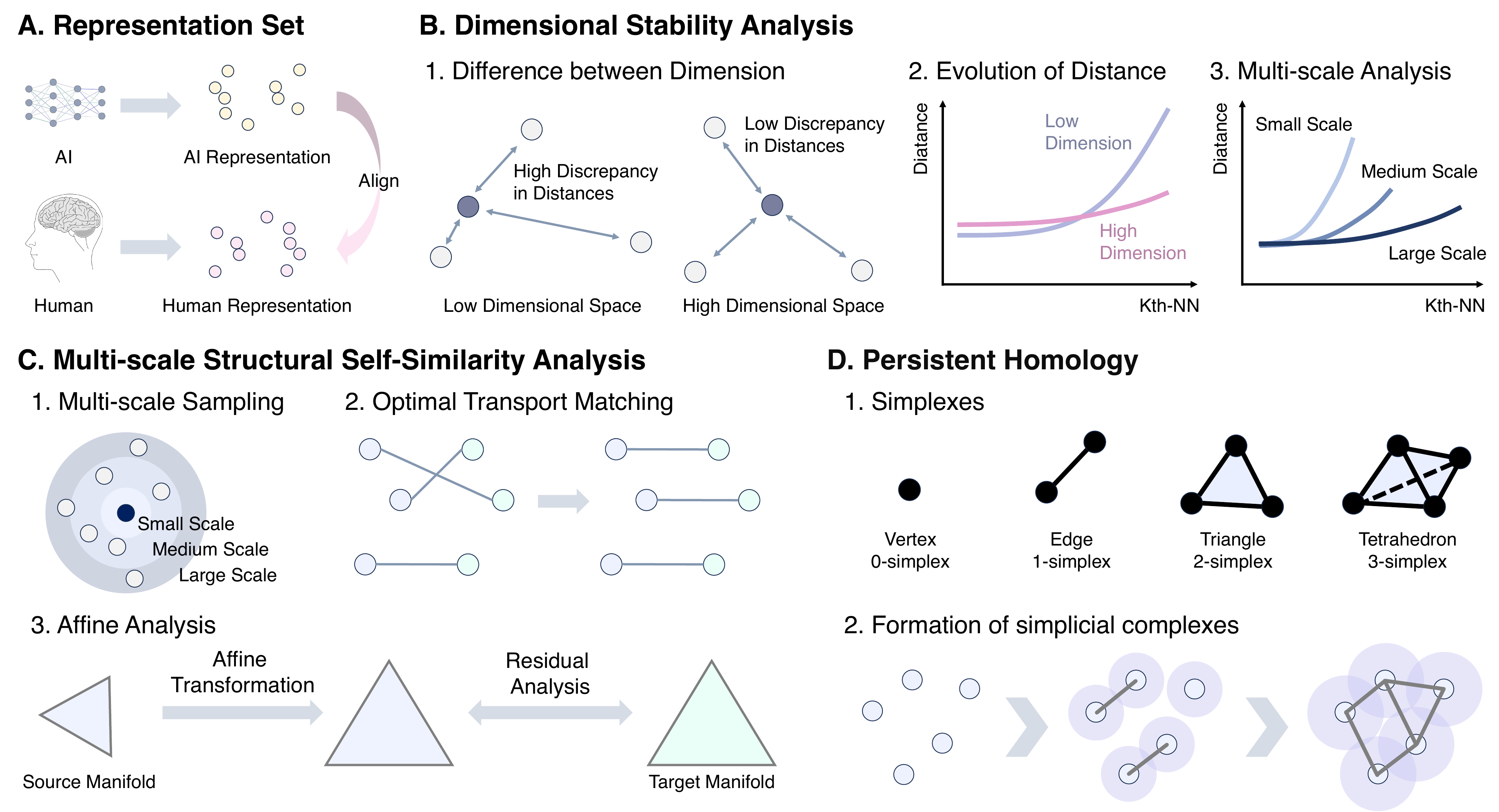}
  \caption{\textbf{Overview of the analytical framework}. \textbf{(A)} Representation sets: The analysis focuses on the representation sets formed by AI and brain responses under image stimuli. \textbf{(B)} Multi-scale dimensional analysis: Dimensionality is estimated by analyzing how distance relationships among samples change from nearest to farthest neighbors across scales. \textbf{(C)} Multi-scale structural analysis: Structural consistency is assessed across varying neighborhood scales to capture structural self-similarity. \textbf{(D)} Persistent homology analysis: Identifying the scales at which structural features emerge most prominently, capturing the significant structural scales in both embeddings and neural manifolds.
}
  \label{fig:method}
\end{figure}

\subsection{Alignment with fMRI Responses}

We use the Natural Scenes Dataset (NSD), a high-resolution fMRI dataset collected at 7T from 8 subjects viewing natural scene images \citep{allen2022massive}. We computed embeddings for 60 different pre-trained vision models and aligned these embeddings with fMRI data using ridge regression (Figure \ref{fig:method} A). To assess the alignment performance, we use the coefficient of determination ($R^2$), which serves as the \textbf{alignment score} throughout this study. Implementation details of the fMRI alignment procedure are provided in Appendix \ref{supple: fMRI Alignment}, and the specific configurations of the 60 models are detailed in Appendix \ref{supple: model details}.

\subsection{Multi-Scale Analysis of Embedding Geometry}

To quantify scale-invariance, we assess two structural properties: dimensional stability and multi-scale self-similarity. Dimensional stability measures the consistency of structural complexity across scales using Maximum Likelihood Estimator (MLE) \citep{levina2004maximum} (Figure \ref{fig:method} B). Multi-scale self-similarity assesses the preservation of local structural patterns under scale changes via affine transformations (Figure \ref{fig:method} C).

\paragraph{Dimensional stability.} 

From low-dimensional to high-dimensional spaces, the differences in distances between points gradually diminish, allowing us to leverage this phenomenon to estimate the dimensionality of the underlying manifold based on the samples (Figure \ref{fig:method} B) .

For a given point \( z \in \mathbb{R}^d \), we compute the dimension based on its \( k \) nearest neighbors:

\begin{equation}
\hat{m}_k(z) = \left[ \frac{1}{k - 1} \sum_{j=1}^{k - 1} \log \frac{T_k(z)}{T_j(z)} \right]^{-1}
\end{equation}

where \( T_k(z) \) denotes the Euclidean distance between the point \( z \) and its \( k \)-th nearest neighbor.

Averaging over all points yields the mean dimension at scale \( k \). By adjusting \( k \), we capture local to global structures, with smaller \( k \) focusing on finer local details and larger \( k \) revealing broader structural patterns.

Dimension stability is assessed by the slope of dimension values across increasing \( k \), indicating how consistently the structural complexity is maintained across scales.  

Furthermore, given that lower dimensionality in deep learning model embeddings is often considered desirable, we also examine the mean dimension across different \( k \) values to explore whether the average dimensionality can serve as a predictor of alignment performance. More details are provided in Appendix \ref{supple: details of dimension}.

\paragraph{Multi-Scale Structural Self-Similarity.}

To assess structural consistency across scales, we examine self-similarity based on affine transformations. Two point clouds \( X, Y \subset \mathbb{R}^d \) are considered affinely equivalent if there exists a transformation that preserves relative spatial configurations under scaling, rotation, and translation.



We assess structural self-similarity through a three-step approach (Figure \ref{fig:method} C):

\textbf{(1) Multi-scale Sampling.}  
This step extracts samples representing different scales of local structure.  

For a randomly selected anchor point \( z^* \), we define neighborhood sizes \( K_i \in \{C_1, C_2, \ldots, C_n\} \), where \( C_1 < C_2 < \ldots < C_n \), and identify the \( K_i \) nearest neighbors. From each set of neighbors, we randomly sample 1000 points to construct point clouds \( X_{K_i} \), ensuring a consistent sample size across scales. Each sampled set reflects a distinct structural scale while maintaining the same number of samples (Implementation details are provided in Appendix \ref{supple: Sampling Strategy}). 

\textbf{(2) Optimal Transport Matching.}  
To align sampled neighborhoods at different scales, we compute the optimal transport plan that minimizes transport cost, ensuring structural correspondences are preserved for subsequent alignment. We also provide details of optimal transport matching and a comparative analysis in Appendix \ref{supple: Optimal Transport} to evaluate the impact of incorporating optimal transport matching versus not applying it.

\textbf{(3) Affine Alignment.}  
After establishing correspondences, we apply affine transformations to minimize residuals between matched points. The key metrics are the average affine residual, reflecting overall structural similarity, and the affine residual slope, indicating how stability changes across scales.

\subsection{Quantifying Structural Scale Distribution via Persistent Homology}

Persistent homology provides a framework for quantifying the distribution of topological features across scales, capturing both the emergence and persistence of structural patterns. We assess structural distribution by identifying critical scales through the birth and death scales of 1-dimensional homology (\( H_1 \)) and evaluating structural similarity using the Jensen–Shannon (JS) divergence between birth-death distributions in AI embeddings and neural manifolds. A lower JS divergence suggests greater similarity in structural patterns across scales. To balance computational feasibility, we focus on \( H_1 \), as higher-dimensional homology (\( H_k, k \geq 2 \)) substantially increases complexity. Implementation details are provided in the Appendix \ref{supple: persistent homology}.

\section{Results}

\subsection{Alignment Performance Across Regions}

We evaluated alignment performance using embeddings generated from 60 different pretrained models, aligning them with fMRI responses across visual and visual association areas in the posterior cortex. 

\begin{figure}[htbp]
  \centering
  \includegraphics[width=0.99\linewidth]{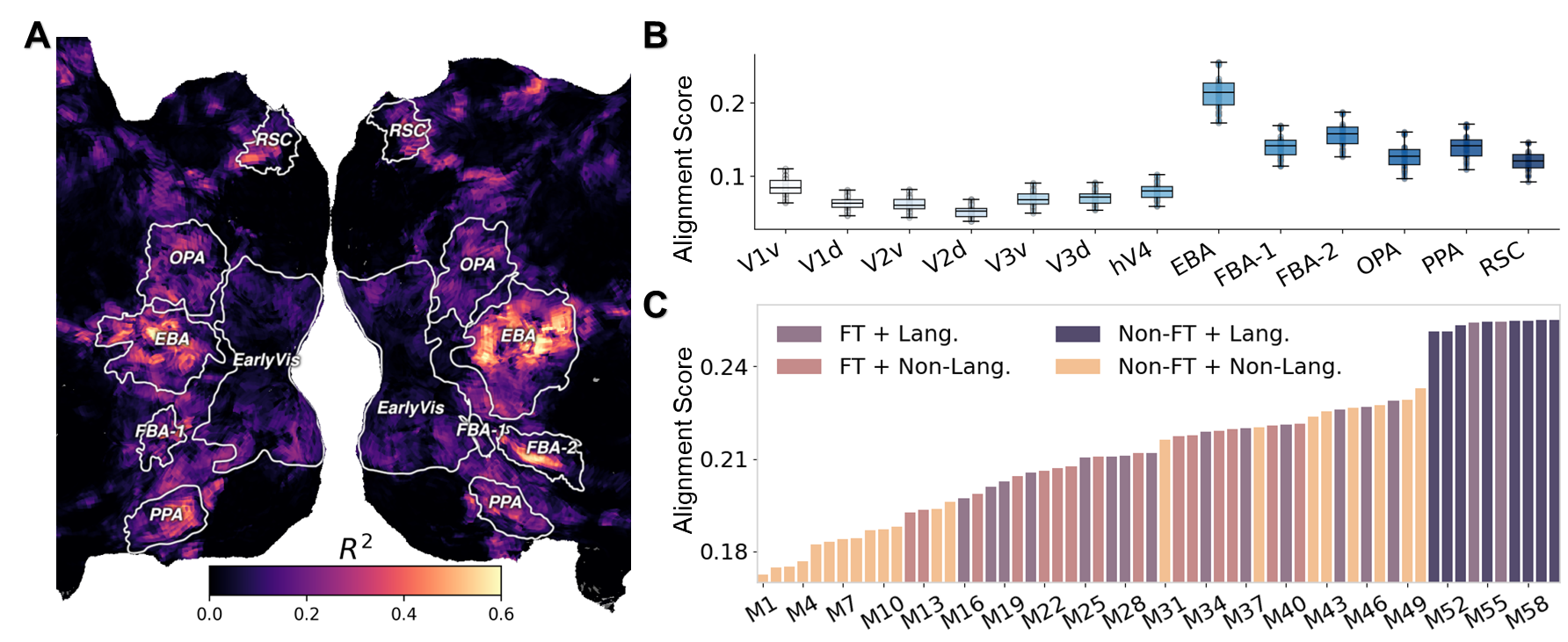}
  \caption{\textbf{Alignment performance across brain regions and models}. \textbf{(A)} Visualizatio of alignment performance across regions, with higher alignment observed in higher-level visual cortex. \textbf{(B)} Comparison of alignment performance across models and regions. Each box represents the distribution of alignment scores for multiple models within a specific brain region. \textbf{(C)} Alignment performance in EBA for all models, grouped by pretraining modality (with/without language) and fine-tuning status. Models pretrained with language and without fine-tuning exhibit significantly stronger alignment than other model groups.}
  \label{fig:alignment}
\end{figure}

The results show that higher-level regions, particularly the extrastriate body area (EBA), exhibit stronger alignment compared to early visual areas (V1, V2) (Figure \ref{fig:alignment} A and B). Notably, the alignment performance in EBA varies significantly across models, with alignment score ranging from 0.16 to 0.25, highlighting substantial differences in model effectiveness even within the same cortical region (Figure \ref{fig:alignment} C).

When grouping models based on whether they were fine-tuned and whether their pretraining phase included a language modality, we find that embeddings from models without fine-tuning but pretrained with a language modality significantly outperform others in alignment (The dark purple histogram in Figure \ref{fig:alignment} C).

Detailed alignment results across models, regions of interest (ROIs), and subjects are provided in Appendix \ref{supple: AI-Brain Alignment}.

\subsection{Multi-Scale Dimensional Stability Predicts Neural Alignment}

\begin{figure}[htbp]
  \centering
  \includegraphics[width=0.95\linewidth]{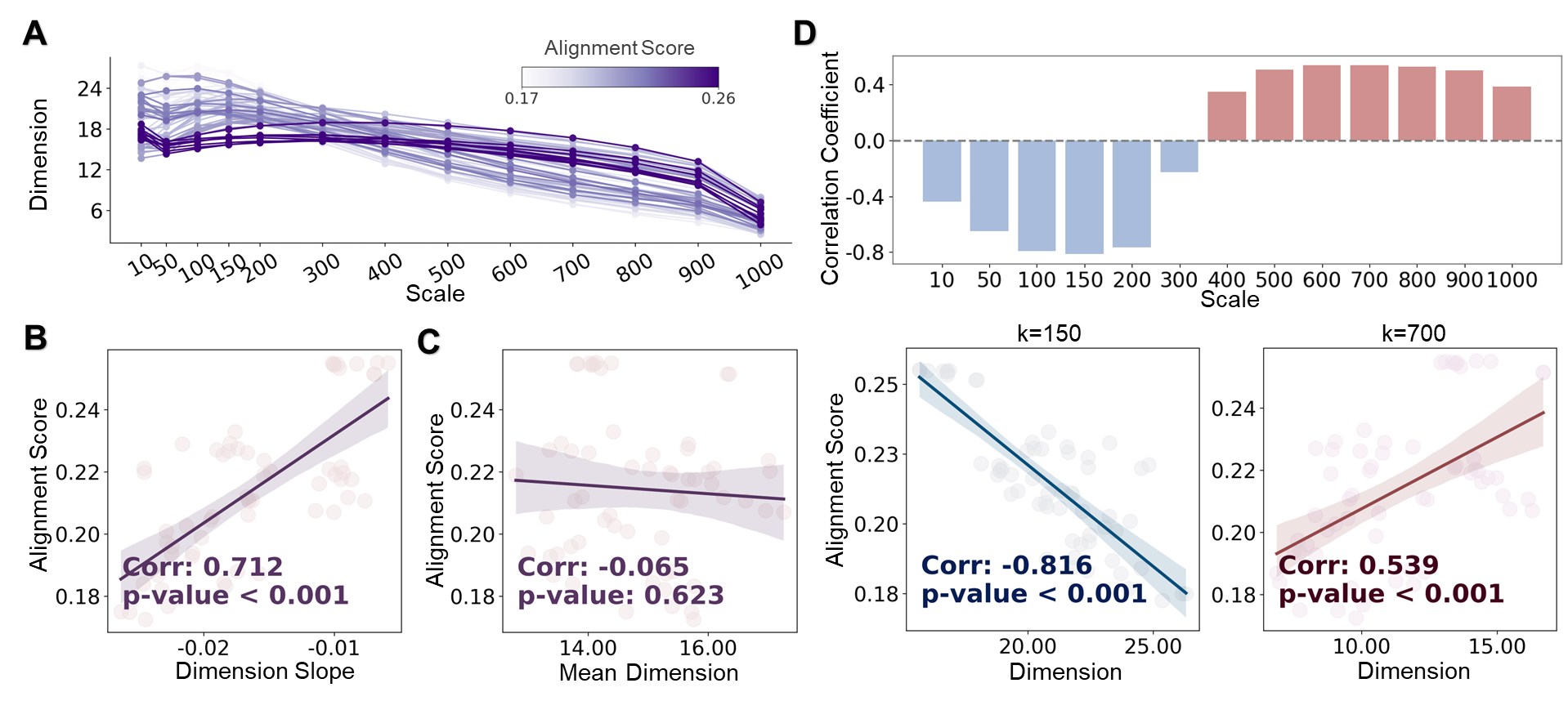}
  \caption{\textbf{Analysis of dimensional stability and alignment performance} \textbf{(A)} Dimensionality across scales for all models, with darker colors indicating better alignment performance. \textbf{(B)} Models with flatter slope exhibit stronger alignment. \textbf{(C)} No significant correlation between mean dimensionality across scales and alignment performance. \textbf{(D)} Scale-specific analysis of dimensionality and alignment, revealing contrasting trends: lower dimensions at smaller scales correlate with better alignment, while higher dimensions at larger scales are associated with stronger alignment.}
  \label{fig:dimensionResult}
\end{figure}

We examined whether the scale-invariance properties of embeddings can predict alignment quality by analyzing how their dimensionality varies across scales. Overall, we observed considerable variability in dimensional stability across scales (Figure \ref{fig:dimensionResult} A), with embeddings exhibiting more consistent dimensions across scales (i.e., flatter slopes) achieving stronger alignment with fMRI responses (Figure \ref{fig:dimensionResult} B) . 

In contrast, the mean dimension across scales showed no significant correlation with alignment, indicating that overall complexity alone is not predictive of alignment quality (Figure \ref{fig:dimensionResult} C).

At specific scales, the relationship between dimensionality and alignment shifted (Figure \ref{fig:dimensionResult} D). At smaller scale, lower dimensions were associated with higher alignment scores, suggesting that more compact local structures are more brain-like. In this range, the negative correlation between dimension and alignment reached as high as \( -0.816 \). Conversely, at larger scales, the relationship reversed, with higher dimensions predicting better alignment, indicating that more complex global structures become advantageous at broader resolutions.

These findings suggest that while average dimensionality is not predictive of alignment, the slope of the dimension curve across scales provides a more robust indicator of how well embeddings align with neural representations. Detailed results for other ROIs are provided in Appendix~\ref{supple: Dimensionality}.

\subsection{Multi-Scale Structural Similarity Predicts Neural Alignment}



In this section, we assessed the structural similarity between embeddings at the smallest scale and those at larger scales.

\begin{figure}[htbp]
  \centering
  \includegraphics[width=0.95\linewidth]{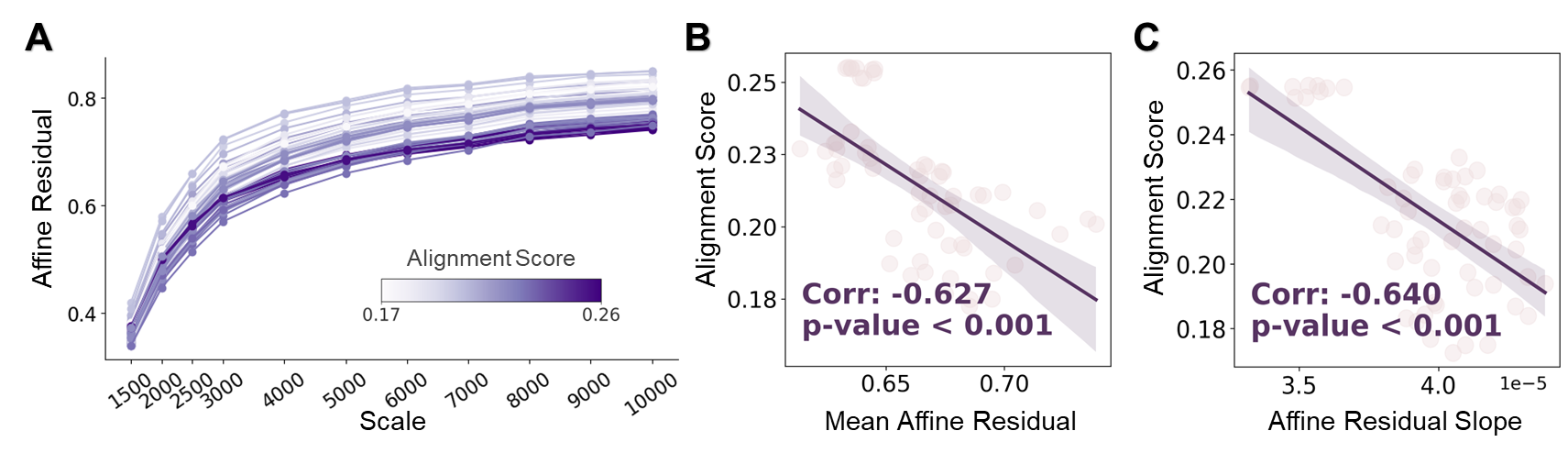}
  \caption{\textbf{Analysis of structural self-similarity and alignment performance.} \textbf{(A)} Affine residuals between the smallest-scale structure and larger-scale structures for different model embeddings, with darker colors indicating better alignment. \textbf{(B)} Lower residuals (i.e., higher structural similarity) are associated with stronger alignment. \textbf{(C)} Models with more stable affine residuals across scales (i.e., flatter slope) exhibit better alignment.}
  \label{fig:selfSimilarity}
\end{figure}

Figure \ref{fig:selfSimilarity} A shows the affine residuals across scales, with the best-aligned models maintaining the lowest residuals consistently. Figures \ref{fig:selfSimilarity} B and C further illustrate that lower average affine residuals and flatter residual slopes are both associated with higher alignment scores, indicating that structural self-similarity across scales effectively predicts alignment performance.

Supplementary results, including detailed analyses of affine residual metrics across models, ROIs, subjects, and different scales are provided in Appendix \ref{supple: Structural Self-Similarity}.

\subsection{Analysis of Structural Distribution Scales of Embeddings and fMRI through Persistent Homology}

We applied persistent homology to quantify the structural distribution of \( H_1 \) features in both embeddings and fMRI data, focusing on voxels extracted from EBA. 

\begin{figure}[htbp]
  \centering
  \includegraphics[width=0.95\linewidth]{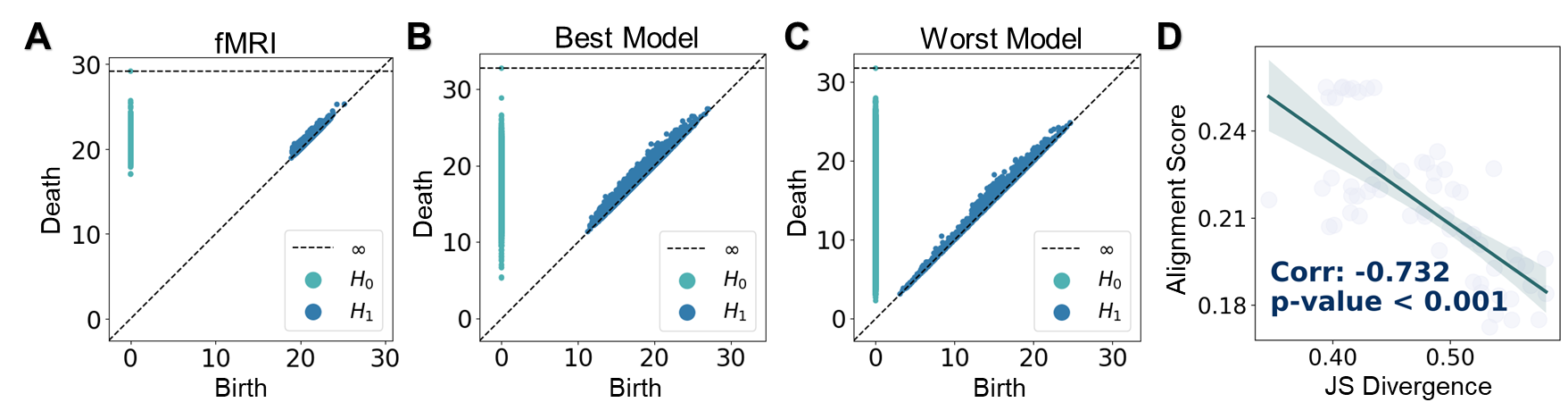}
  \caption{\textbf{Analysis of topological feature distribution across scales.} \textbf{(A)} Distribution of \( H_1 \) scales in EBA, \textbf{(B)} the best-aligned model, and \textbf{(C)} the worst-aligned model. EBA shows a concentrated distribution, the best model shows a moderately dispersed distribution, and the worst model displays the most dispersed pattern. (D) JS divergence quantifying the similarity in topological scale distributions, confirming that embeddings with more compact scale distributions align better with fMRI data.
}
  \label{fig:structuralDistribution}
\end{figure}


As shown in Figure \ref{fig:structuralDistribution} A, \( H_1 \) feature distribution in EBA reveals a distinct pattern of compact structural scales. The birth scales are densely concentrated within a narrow range, indicating that neural representations prioritize a focused set of structural scales for more efficient encoding.

Next, we compared the persistent diagrams of embeddings from the best and worst models in terms of alignment with EBA. The best-performing model exhibited a similarly compact distribution of \( H_1 \) features, closely mirroring the concentrated pattern observed in EBA (Figure \ref{fig:structuralDistribution} B). In contrast, the worst-performing model displayed a more diffuse and broadly distributed set of structural scales, suggesting a less focused structural organization (Figure \ref{fig:structuralDistribution} C).

To further quantify the relationship between structural distribution and alignment, we computed the JS divergence between the birth-death distributions of \( H_1 \) features in embeddings and EBA fMRI data. Figure \ref{fig:structuralDistribution} D revealed a strong negative correlation between JS divergence and alignment scores, indicating that models with more compact and tightly clustered structural distributions align more effectively with neural data.

Additional results across ROIs and subjects are presented in Appendix \ref{supple: Persistent Homology Results}.

\subsection{Influence of Data Volume, Language Modality and Label Supervision on Multi-Scale Embedding Structure}

To assess how external world information influences the formation of scale-invariant structures in embeddings, we examined the effects of data volume, language modality, and label supervision. These factors encode different aspects of external information, potentially shaping the scale-invariant properties of embeddings. We grouped models based on these factors and analyzed their dimensional stability and structural similarity across scales.

\begin{figure}[htbp]
  \centering
  \includegraphics[width=0.98\linewidth]{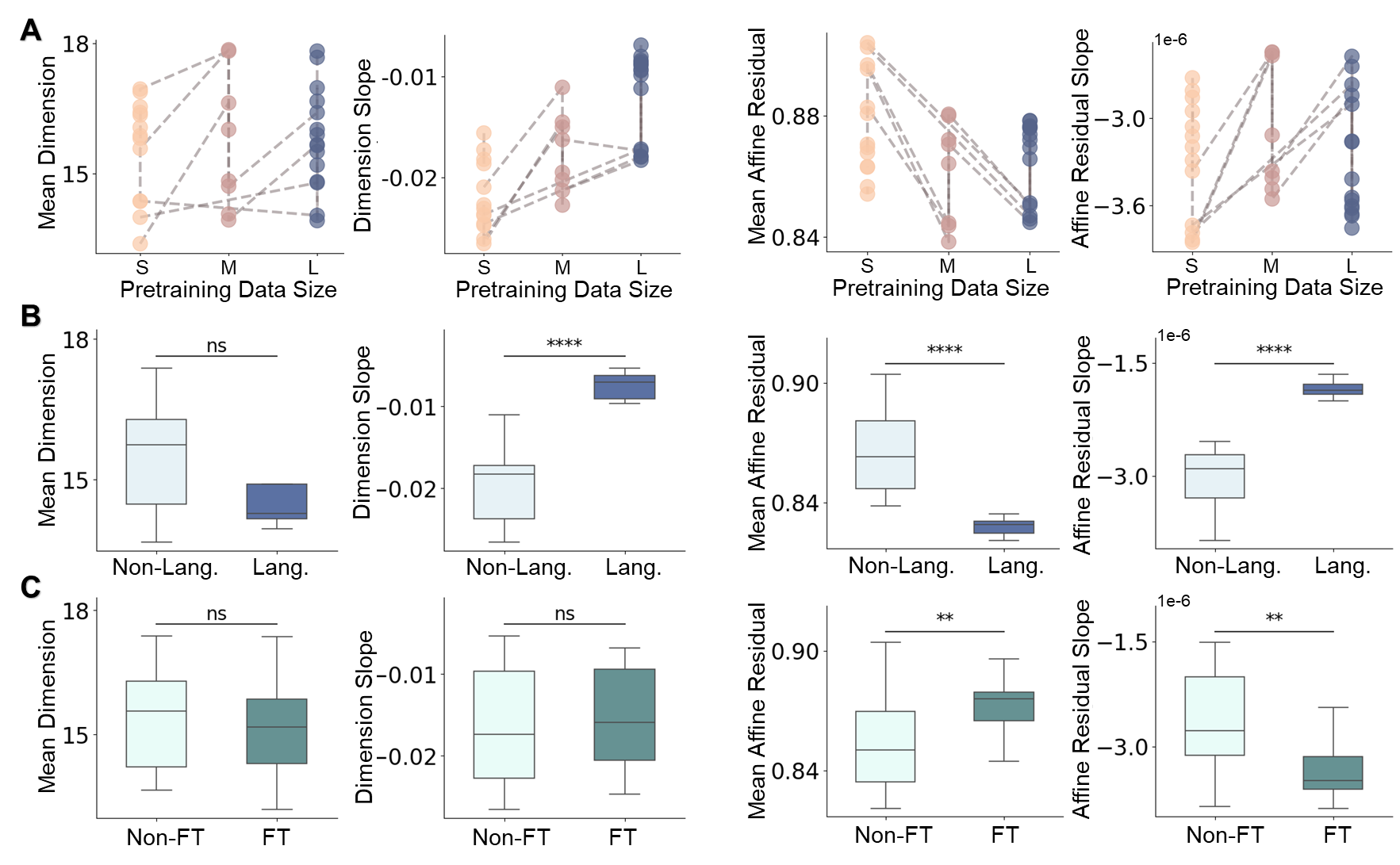}
  \caption{\textbf{Effects of data scale, modality, and fine-tuning on embedding structure.} \textbf{(A)} Larger datasets and \textbf{(B)} the inclusion of language modality enhance dimensional stability and structural self-similarity across scales. \textbf{(C)} Fine-tuning disrupts scale-invariance, primarily by weakening structural self-similarity.
}
  \label{fig:influence}
\end{figure}

\paragraph{Data Volume and Language Modality Promote Structural Consistency}  
Models trained on larger datasets exhibited higher average dimensions, flatter dimensional slopes, and greater structural self-similarity across scales (Figure \ref{fig:influence} A). This suggests that increased data volume not only enriches structural information but also fosters more coherent multi-scale organization. Language modality further amplified structural self-similarity (Figure \ref{fig:influence} B). Vision-language models showed flatter dimensional slopes and stronger cross-scale consistency, indicating that language may introduce multi-scale semantic structures that stabilize embedding organization.

\paragraph{Label Supervision Disrupts Scale-Invariant Structures}  
In contrast, although label fine-tuning does not explicitly alter dimensional stability, it reduces cross-scale structural similarity, suggesting that task-specific supervision partially disrupts the multi-scale coherence established during pretraining (Figure \ref{fig:influence} C).

Overall, larger multimodal datasets promote more scale-invariant structures, whereas label supervision introduces task-specific constraints that disrupt structural consistency.

\section{Conclusion} 

Our findings demonstrate that embedding manifolds with \textbf{compact, scale-invariant structures} align more effectively with fMRI activity in the visual cortex. This supports the hypothesis that external structural constraints, particularly scale-invariance, shape both artificial and neural representations. 

In addition to analyzing embeddings, we also assessed the scale-invariance of fMRI data (Appendix \ref{supple: fMRI scale-variance}). Our findings indicate that fMRI data exhibits stronger structural self-similarity across scales compared to AI embeddings, maintaining consistent organization despite scale variations. At smaller scales, the dimensionality of fMRI data rises rapidly, likely due to structural degradation that amplifies local variations. At larger scales, the dimensionality stabilizes and remains lower than that of AI embeddings, further suggesting that neural manifolds are structurally organized within specific scales, concentrating information more effectively at certain levels of granularity.

Pretraining on large, multimodal datasets promotes more compact, scale-invariant structures, whereas fine-tuning shifts focus to task-specific patterns, disrupting this organization. This trade-off highlights the tension between task adaptation and preserving structural alignment with neural data.

Overall, our results suggest that compact, scale-invariant structures are crucial for aligning artificial and neural representations, with neural manifolds concentrating structural features at specific scales while maintaining self-similarity.

\section{Limitations and Future Work} 

A key limitation is the difficulty in \textbf{controlling multiple model variables simultaneously}, such as dataset size, modality, and fine-tuning, making it challenging to isolate their individual effects. Future work could address this by systematically varying these factors in controlled settings.

Moreover, neural systems represent information in a distributed manner across multiple brain regions \citep{rissman2012distributed, small1995distributed}, whereas our analysis focuses on single region. Extending this framework to multiple brain regions while minimizing the influence of irrelevant information is both a challenge and a critical direction for advancing neural representation analysis.

\bibliography{neurips_2025}


\appendix

\newpage

\section{Details of fMRI Alignment}
\label{supple: fMRI Alignment}

\paragraph{Embedding extraction.} 
For each image stimulus, we extract visual embeddings from pretrained vision models. All embeddings are projected to 300 dimensions using principal component analysis (PCA) for standardization across models and computational efficiency.

\paragraph{Voxel-wise encoding model.} 
Let $z_i \in \mathbb{R}^{300}$ be the embedding of image $i$, and let $r_{ij}$ be the fMRI response of voxel $j$ to image $i$. We train a voxel-specific ridge regression model:

\begin{equation}
r_{ij} \approx \mathbf{w}_j^\top \mathbf{z}_i + \epsilon.
\end{equation}

Model fitting is done using 5-fold cross-validation on an 80/20 training/test split. The regularization parameter is tuned via nested cross-validation.

\paragraph{Evaluation metrics.} 
Alignment quality is quantified using both Pearson correlation and coefficient of determination ($R^2$) on held-out data.

\section{Model Architecture and Implementation Details}
\label{supple: model details}

In this study, we employed the ConvNeXt architecture for generating embeddings across multiple models and datasets. ConvNeXt is a family of pure convolutional neural networks designed as a modernized version of ResNet, incorporating architectural elements inspired by Vision Transformers while maintaining the simplicity and computational efficiency of traditional ConvNets \citep{liu2022convnet}. 

ConvNeXt is constructed by progressively adapting the ResNet architecture to incorporate design principles from Vision Transformers, such as large kernel sizes, depthwise convolutions, and layer normalization. The architecture consists of four main stages with varying resolutions, each composed of multiple residual blocks. Key modifications in ConvNeXt include the replacement of ReLU with GELU activation for smoother gradients, the adoption of large kernel sizes (e.g., 7x7) to increase receptive fields, the implementation of depthwise separable convolutions for efficient spatial mixing, the utilization of Layer Normalization instead of Batch Normalization, and the employment of a patchify stem to align with Vision Transformer preprocessing. ConvNeXt achieves competitive performance across multiple visual recognition tasks, surpassing the Swin Transformer in terms of accuracy and scalability while maintaining computational efficiency.

\paragraph{Implementation via Huggingface API:}  
For this analysis, we leveraged the Huggingface Transformers library to access ConvNeXt models pre-trained on various datasets. Huggingface provides several ConvNeXt variants pre-trained on datasets such as ImageNet-1K and ImageNet-22K, allowing us to systematically compare model embeddings derived from different pre-training datasets while controlling for architectural variations.

\paragraph{Computational Resources}  
We utilized an NVIDIA GeForce RTX 3080 GPU with 10GB GDDR6X memory, a 320-bit memory interface, and a memory bandwidth of 760 GB/s for embedding extraction using the Huggingface Transformers API. This setup adequately meets the computational requirements without requiring additional resources.

\paragraph{Rationale for Using ConvNeXt:}  
The primary motivation for utilizing ConvNeXt in this study is its modular design, enabling consistent model architecture across multiple pre-training datasets. This setup allows for a controlled analysis of how pre-training data impacts alignment with neural data while minimizing confounding effects from architectural changes. Moreover, ConvNeXt’s architectural simplicity and computational efficiency facilitate the generation of embeddings across multiple participants and brain regions, ensuring robust comparative analyses.

In summary, the ConvNeXt architecture serves as a robust and scalable backbone for examining the influence of pre-training datasets on model-brain alignment, providing a controlled framework for evaluating structural similarities between model embeddings and neural data.

\begin{table}[H]
  \caption{Model information including model index (sorted by alignment score in EBA), pretraining dataset, fine-tuning, and language modality usage.}
  \label{model-info-table}
  \centering
  \resizebox{\textwidth}{!}{%
  \begin{tabular}{lllll}
    \toprule
    \textbf{Model Name} & \textbf{Index} & \textbf{Pretraining Dataset} & \textbf{FT} & \textbf{Lang} \\
    \midrule
    \texttt{convnext\_nano.d1h\_in1k} & M1 & ImageNet-1K & \ding{55} & \ding{55} \\
    \texttt{convnext\_nano\_ols.d1h\_in1k} & M2 & ImageNet-1K & \ding{55} & \ding{55} \\
    \texttt{convnext\_tiny\_hnf.a2h\_in1k} & M3 & ImageNet-1K & \ding{55} & \ding{55} \\
    \texttt{convnext\_pico\_ols.d1\_in1k} & M4 & ImageNet-1K & \ding{55} & \ding{55} \\
    \texttt{convnext\_pico.d1\_in1k} & M5 & ImageNet-1K & \ding{55} & \ding{55} \\
    \texttt{convnext\_atto\_ols.a2\_in1k} & M6 & ImageNet-1K & \ding{55} & \ding{55} \\
    \texttt{convnext\_large.fb\_in1k} & M7 & ImageNet-1K & \ding{55} & \ding{55} \\
    \texttt{convnext\_femto.d1\_in1k} & M8 & ImageNet-1K & \ding{55} & \ding{55} \\
    \texttt{convnext\_base.fb\_in1k} & M9 & ImageNet-1K & \ding{55} & \ding{55} \\
    \texttt{convnext\_atto.d2\_in1k} & M10 & ImageNet-1K & \ding{55} & \ding{55} \\
    \texttt{convnext\_femto\_ols.d1\_in1k} & M11 & ImageNet-1K & \ding{55} & \ding{55} \\
    \texttt{convnext\_small.in12k\_ft\_in1k\_384} & M12 & ImageNet-12K & \ding{51} & \ding{55} \\
    \texttt{convnext\_small.in12k\_ft\_in1k} & M13 & ImageNet-12K & \ding{51} & \ding{55} \\
    \texttt{convnext\_small.fb\_in1k} & M14 & ImageNet-1K & \ding{55} & \ding{55} \\
    \texttt{convnext\_tiny.fb\_in1k} & M15 & ImageNet-1K & \ding{55} & \ding{55} \\
    \texttt{convnext\_base.clip\_laion2b\_augreg\_ft\_in1k} & M16 & LAION-2B & \ding{51} & \ding{51} \\
    \texttt{convnext\_nano.in12k\_ft\_in1k} & M17 & ImageNet-12K & \ding{51} & \ding{55} \\
    \texttt{convnext\_large\_mlp.clip\_laion2b\_augreg\_ft\_in1k} & M18 & LAION-2B & \ding{51} & \ding{51} \\
    \texttt{convnext\_large\_mlp.clip\_laion2b\_augreg\_ft\_in1k\_384} & M19 & LAION-2B & \ding{51} & \ding{51} \\
    \texttt{convnext\_tiny.in12k\_ft\_in1k} & M20 & ImageNet-12K & \ding{51} & \ding{55} \\
    \texttt{convnext\_base.clip\_laiona\_augreg\_ft\_in1k\_384} & M21 & LAION-A & \ding{51} & \ding{51} \\
    \texttt{convnext\_tiny.in12k\_ft\_in1k\_384} & M22 & ImageNet-12K & \ding{51} & \ding{55} \\
    \texttt{convnext\_tiny.fb\_in22k\_ft\_in1k\_384} & M23 & ImageNet-22K & \ding{51} & \ding{55} \\
    \texttt{convnext\_tiny.fb\_in22k\_ft\_in1k} & M24 & ImageNet-22K & \ding{51} & \ding{55} \\
    \texttt{convnext\_large\_mlp.clip\_laion2b\_soup\_ft\_in12k\_in1k\_384} & M25 & LAION-2B & \ding{51} & \ding{51} \\
    \texttt{convnext\_small.fb\_in22k\_ft\_in1k\_384} & M26 & ImageNet-22K & \ding{51} & \ding{55} \\
    \texttt{convnext\_base.clip\_laion2b\_augreg\_ft\_in12k\_in1k} & M27 & LAION-2B & \ding{51} & \ding{51} \\
    \texttt{convnext\_large\_mlp.clip\_laion2b\_soup\_ft\_in12k\_in1k\_320} & M28 & LAION-2B & \ding{51} & \ding{51} \\
    \texttt{convnext\_small.fb\_in22k\_ft\_in1k} & M29 & ImageNet-22K & \ding{51} & \ding{55} \\
    \texttt{convnext\_base.clip\_laion2b\_augreg\_ft\_in12k\_in1k\_384} & M30 & LAION-2B & \ding{51} & \ding{51} \\
    \texttt{convnext\_small.in12k} & M31 & ImageNet-12K & \ding{55} & \ding{55} \\
    \texttt{convnext\_base.fb\_in22k\_ft\_in1k\_384} & M32 & ImageNet-22K & \ding{51} & \ding{55} \\
    \texttt{convnext\_large.fb\_in22k\_ft\_in1k\_384} & M33 & ImageNet-22K & \ding{51} & \ding{55} \\
    \texttt{convnext\_xxlarge.clip\_laion2b\_soup\_ft\_in1k} & M34 & LAION-2B & \ding{51} & \ding{51} \\
    \texttt{convnext\_base.fb\_in22k\_ft\_in1k} & M35 & ImageNet-22K & \ding{51} & \ding{55} \\
    \texttt{convnext\_xlarge.fb\_in22k\_ft\_in1k\_384} & M36 & ImageNet-22K & \ding{51} & \ding{55} \\
    \texttt{convnext\_large\_mlp.clip\_laion2b\_soup\_ft\_in12k\_384} & M37 & LAION-2B & \ding{51} & \ding{51} \\
    \texttt{convnext\_tiny.in12k} & M38 & ImageNet-12K & \ding{55} & \ding{55} \\
    \texttt{convnext\_large.fb\_in22k\_ft\_in1k} & M39 & ImageNet-22K & \ding{51} & \ding{55} \\
    \texttt{convnext\_large\_mlp.clip\_laion2b\_augreg\_ft\_in12k\_384} & M40 & LAION-2B & \ding{51} & \ding{51} \\
    \texttt{convnext\_xlarge.fb\_in22k\_ft\_in1k} & M41 & ImageNet-22K & \ding{51} & \ding{55} \\
    \texttt{convnext\_nano.in12k} & M42 & ImageNet-12K & \ding{55} & \ding{55} \\
    \texttt{convnext\_small.fb\_in22k} & M43 & ImageNet-22K & \ding{55} & \ding{55} \\
    \texttt{convnext\_base.clip\_laion2b\_augreg\_ft\_in12k} & M44 & LAION-2B & \ding{51} & \ding{51} \\
    \texttt{convnext\_large.fb\_in22k} & M45 & ImageNet-22K & \ding{55} & \ding{55} \\
    \texttt{convnext\_xxlarge.clip\_laion2b\_soup\_ft\_in12k} & M46 & LAION-2B & \ding{51} & \ding{51} \\
    \texttt{convnext\_tiny.fb\_in22k} & M47 & ImageNet-22K & \ding{55} & \ding{55} \\
    \texttt{convnext\_large\_mlp.clip\_laion2b\_soup\_ft\_in12k\_320} & M48 & LAION-2B & \ding{51} & \ding{51} \\
    \texttt{convnext\_xlarge.fb\_in22k} & M49 & ImageNet-22K & \ding{55} & \ding{55} \\
    \texttt{convnext\_base.fb\_in22k} & M50 & ImageNet-22K & \ding{55} & \ding{55} \\
    \texttt{convnext\_xxlarge.clip\_laion2b\_rewind} & M51 & LAION-2B & \ding{55} & \ding{51} \\
    \texttt{convnext\_xxlarge.clip\_laion2b\_soup} & M52 & LAION-2B & \ding{55} & \ding{51} \\
    \texttt{convnext\_large\_mlp.clip\_laion2b\_augreg} & M53 & LAION-2B & \ding{55} & \ding{51} \\
    \texttt{convnext\_large\_mlp.clip\_laion2b\_ft\_320} & M54 & LAION-2B & \ding{51} & \ding{51} \\
    \texttt{convnext\_base.clip\_laiona} & M55 & LAION-A & \ding{55} & \ding{51} \\
    \texttt{convnext\_large\_mlp.clip\_laion2b\_ft\_soup\_320} & M56 & LAION-2B & \ding{51} & \ding{51} \\
    \texttt{convnext\_base.clip\_laion2b} & M57 & LAION-2B & \ding{55} & \ding{51} \\
    \texttt{convnext\_base.clip\_laiona\_320} & M58 & LAION-A & \ding{55} & \ding{51} \\
    \texttt{convnext\_base.clip\_laion2b\_augreg} & M59 & LAION-2B & \ding{55} & \ding{51} \\
    \texttt{convnext\_base.clip\_laiona\_augreg\_320} & M60 & LAION-A & \ding{55} & \ding{51} \\
    \bottomrule
    \end{tabular}
  }
\end{table}

\section{Background on Dimensional Stability Analysis}
\label{supple: details of dimension}

\subsection{Fractal Dimension and Correlation Dimension}

The concept of fractal dimension provides a robust framework for quantifying the complexity, irregularity and structural intricacies of datasets across different scales. Unlike traditional Euclidean dimensions, fractal dimensions capture the degree to which a set fills space as the observation scale varies. This is particularly useful in the context of high-dimensional data and manifold learning.

\begin{definition}
\textbf{Fractal Dimension}: The fractal dimension $D_f$ is defined as the scaling exponent that quantifies how the number of self-similar structures in a set changes with the scale of observation. Mathematically, it can be expressed as:
\begin{equation}
D_f = \lim_{\epsilon \to 0} \frac{\log N(\epsilon)}{\log (1/\epsilon)},
\end{equation}
where $N(\epsilon)$ represents the number of $\epsilon$-sized boxes required to cover the set.

A higher fractal dimension indicates a more intricate structure with greater self-similarity across multiple scales.

\end{definition}

\subsubsection{Correlation Dimension}

The correlation dimension is a specific type of fractal dimension that focuses on the spatial distribution properties of a point set, effectively capturing statistical sparsity and clustering behavior. It provides a finer-grained analysis of the dataset structure, especially in cases where data points are non-uniformly distributed.

\begin{definition}
\textbf{Correlation Dimension}: The correlation dimension $D_C$ is defined based on the correlation function $C(\epsilon)$ as:
\begin{equation}
D_C = \lim_{\epsilon \to 0} \frac{\log C(\epsilon)}{\log \epsilon},
\end{equation}
where the correlation function $C(\epsilon)$ is given by:
\begin{equation}
C(\epsilon) = \frac{1}{N(N-1)} \sum_{i=1}^N \sum_{j \neq i} \mathbb{I}(\|x_i - x_j\| < \epsilon).
\end{equation}

\begin{itemize}
    \item $N$ is the number of data points.
    \item $\|x_i - x_j\|$ represents the Euclidean distance between points $x_i$ and $x_j$.
    \item $\mathbb{I}(\cdot)$ is the indicator function, equal to $1$ if the condition inside is true and $0$ otherwise.
\end{itemize}

The correlation dimension effectively estimates the likelihood that two randomly chosen points from the dataset are within a distance $\epsilon$ of each other, thus providing insights into the spatial organization of the data.

\end{definition}

\subsection{Fractal Dimension Estimation Using Maximum Likelihood}

To accurately estimate the fractal dimension, particularly the correlation dimension, we employ the Maximum Likelihood Estimation (MLE) method as proposed by Levina and Bickel \citep{levina2004maximum}. This method leverages distances between neighboring points to infer the intrinsic dimensionality of the underlying manifold.

\subsubsection{Fractal Dimension Estimation via MLE}

The MLE-based estimation method utilizes the $k$-nearest neighbor distances to quantify the dimensionality of a dataset. For a given data point $x_i$, the estimated fractal dimension $\hat{m_k}(x_i)$ at a specific $k$ value is defined as:

\begin{equation}
\hat{m_k}(x_i) = \left[\frac{1}{k-1} \sum_{j=1}^{k-1} \log \frac{T_k(x_i)}{T_j(x_i)}\right]^{-1},
\end{equation}
where:

\begin{itemize}
    \item $T_j(x_i)$ is the Euclidean distance from $x_i$ to its $j^{\text{th}}$ nearest neighbor.
    \item $k$ is the number of nearest neighbors considered for the estimation.
\end{itemize}

A smaller $k$ captures local structural variations, emphasizing finer-scale features, while a larger $k$ reflects broader, more global patterns in the data distribution.

\subsubsection{Averaging Over Data Points}

To obtain a robust estimate of the fractal dimension for the entire dataset, we average the estimated dimensions across all data points:

\begin{equation}
\bar{m_k} = \frac{1}{N} \sum_{i=1}^N \hat{m_k}(x_i),
\end{equation}

where $N$ is the total number of data points. This averaging process helps to mitigate the effects of noise and local variations, producing a stable estimate of the fractal dimension at a specific $k$ value.

\subsubsection{Selection of $k$ and Scale Analysis}

The choice of $k$ significantly impacts the resulting dimension estimate:

\begin{itemize}
    \item \textbf{Small $k$}: Focuses on finer-scale structures, capturing local density variations and small-scale clustering.
    \item \textbf{Large $k$}: Emphasizes broader trends, providing a more smoothed, global perspective of the dataset structure.
\end{itemize}

To achieve a comprehensive understanding of the manifold, it is essential to analyze the fractal dimension estimates across multiple $k$ values. This multi-scale analysis helps in identifying how the estimated dimensionality varies with scale, offering deeper insights into the intrinsic complexity of the dataset.

\subsection{Application to Manifold Complexity Analysis}

Estimating the fractal and correlation dimensions enables us to quantify the intrinsic complexity of data manifolds, particularly in high-dimensional settings. By systematically analyzing the scaling behavior and spatial organization of the dataset across different $k$ values, we can effectively characterize manifold structures and their dimensional stability. This approach provides a robust framework for identifying structural patterns, assessing sparsity and detecting non-linear data distributions.

\section{Details of Sampling Strategy}
\label{supple: Sampling Strategy}

In the main text, we described the sampling strategy as “randomly sampling 1000 points.” Here, we provide a more detailed explanation of the sampling procedure, including the selection of anchor points and the implementation of the random sampling process.

\paragraph{Anchor Point Selection:}  
For each analysis, we select three sample points as anchor points. These anchor points are randomly chosen from the dataset, and the final result is obtained by averaging across these three points. This strategy ensures that the analysis captures the structural characteristics of different parts of the data manifold, rather than being biased toward a single anchor point.

\paragraph{Random Sampling Process:}  
At each scale, we perform random sampling without replacement to select 1000 sample points from the neighborhood set. The sampling process is implemented as follows:

\begin{verbatim}
np.random.seed(42)
sampled = nearest_k[np.random.choice(K, 1000, replace=False)]
\end{verbatim}

By setting the random seed to 42, we ensure that the sampling process is consistent across different experimental runs, enabling reproducibility of the results. This approach minimizes sampling variability and ensures that any observed differences in the analysis are attributable to changes in the data or model embeddings rather than stochastic fluctuations in the sampling process.

\section{Details of Optimal Transport Matching}
\label{supple: Optimal Transport}

In this study, we employed Optimal Transport (OT) to align embeddings across different scales, ensuring that structural comparisons are not confounded by spatial misalignments. OT provides a principled framework for matching data points between two sets by minimizing the transportation cost, effectively capturing the optimal correspondence between points.

\paragraph{Optimal Transport Implementation:}  
The implementation of OT was carried out using the Python Optimal Transport (POT) library, which provides a comprehensive set of tools for computing the Earth Mover’s Distance (EMD) and related optimal transport measures. By leveraging the EMD algorithm, we effectively aligned the sample distributions across scales, ensuring that the structural comparisons were based on optimally matched data points.

\begin{figure}[H]
  \centering
  \includegraphics[width=0.95\linewidth]{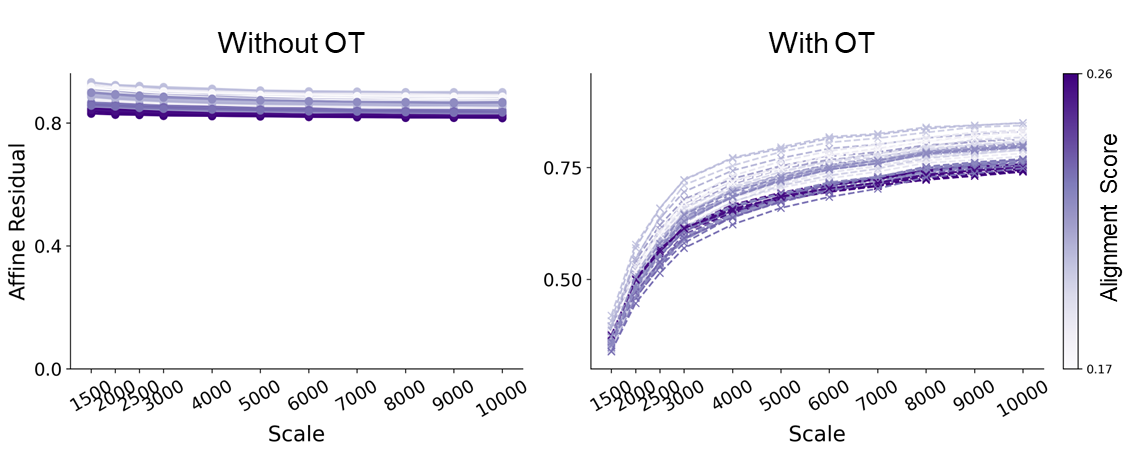}
  \caption{\textbf{Impact of optimal transport on multi-scale structural similarity analysis.} Without optimal transport, affine residuals remain consistently high across scales, obscuring the true structural evolution from the smallest to the largest scale. In contrast, applying optimal transport effectively reduces affine residuals as scales approach the smallest scale, capturing the gradual structural transitions more accurately and reflecting the true multi-scale organization of the manifold.
}
  \label{fig:W_WO_OT_comparison}
\end{figure}

\paragraph{Comparative Analysis: With vs. Without OT}  
We conducted a comparative analysis to evaluate the effect of OT on multi-scale structural similarity analysis, as shown in Figure~\ref{fig:W_WO_OT_comparison}. In both cases, we extracted the embedding at the smallest scale and compared it with embeddings at progressively larger scales to assess structural similarity. Theoretically, as the scales become more similar (i.e., closer to the smallest scale), the structural similarity should increase, resulting in lower affine residuals.

However, without applying OT (left panel), this expected trend is not evident, as affine residuals remain consistently high across scales, failing to capture the expected structural transition. In contrast, after applying OT (right panel), the affine residuals decrease systematically as scales approach the smallest scale, effectively revealing the underlying structural evolution. This demonstrates that OT is essential for accurately capturing the structural consistency across scales, aligning data points more effectively and mitigating spatial discrepancies that obscure true structural patterns.

\section{Background on Persistent Homology Analysis}
\label{supple: persistent homology}

\subsection{Overview of Persistent Homology}

Persistent homology is a robust method in topological data analysis (TDA) that systematically quantifies structural features across multiple scales within a dataset. Unlike traditional geometric analyses that primarily focus on pairwise distances or local neighborhood properties, persistent homology captures higher-order topological structures such as connected components, loops, and voids. By tracking the emergence and persistence of these structures as the scale parameter varies, persistent homology provides a multi-scale representation of the underlying shape of data.

\paragraph{Constructing Simplicial Complexes.}
To perform persistent homology analysis, we first construct a sequence of simplicial complexes from the input data. Given a point cloud \( Z \subset \mathbb{R}^d \), we use the Vietoris–Rips filtration to generate complexes at varying scale parameters \( \varepsilon \). For a specified scale \( \varepsilon \), points whose pairwise distances are less than \( \varepsilon \) are connected to form edges. As the scale increases, higher-dimensional simplices such as triangles and tetrahedrons are iteratively added. This filtration process creates a nested family of simplicial complexes, capturing topological features at multiple scales.

\subsection{Understanding Homology: \( H_0 \), \( H_1 \), and \( H_2 \)}

Homology provides a formal framework for quantifying topological features of different dimensions within a dataset:

\textbf{\( H_0 \) (0-dimensional homology)}: Captures connected components. Each connected cluster of points in the dataset is represented as a separate 0-dimensional feature. As the scale parameter increases, these components may merge, and the disappearance of a component is recorded as its death.

\textbf{\( H_1 \) (1-dimensional homology)}: Captures loop structures. Loops are closed cycles formed by interconnected points that enclose a void or hole. These structures emerge at intermediate scales, reflecting non-trivial connectivity in the dataset. In this study, we focus on \( H_1 \) features as they effectively capture the organization of data points within the manifold, providing insight into intermediate-scale structures and potential clustering patterns.

\textbf{\( H_2 \) (2-dimensional homology)}: Captures voids or cavities. Voids are higher-dimensional holes enclosed by surfaces formed by multiple simplices (e.g., triangles forming a closed surface). Although these features provide additional information about multi-dimensional connectivity, they are less relevant in the context of data embedded in lower-dimensional spaces.

\subsection{Implementing Persistent Homology Analysis Using Ripser and Persim}

In this analysis, we utilize two Python libraries: \textbf{Ripser} and \textbf{Persim}. Ripser is a computationally efficient tool for computing persistent homology, leveraging a cohomological approach for rapid filtration and feature extraction. Persim, on the other hand, provides tools for visualizing persistence diagrams and calculating distances between them. By combining these libraries, we can not only extract topological features efficiently but also systematically compare the structural complexity of different datasets using persistence diagrams.


\section{Supplementary Analysis on AI-Brain Alignment}
\label{supple: AI-Brain Alignment}


To further investigate the voxel-wise alignment analysis presented in the main text, we extended the analysis across all eight participants using multiple models. Here, we focus on two key visualizations that highlight the variability and consistency of alignment performance.

\begin{figure}[H]
  \centering
  \includegraphics[width=0.99\linewidth]{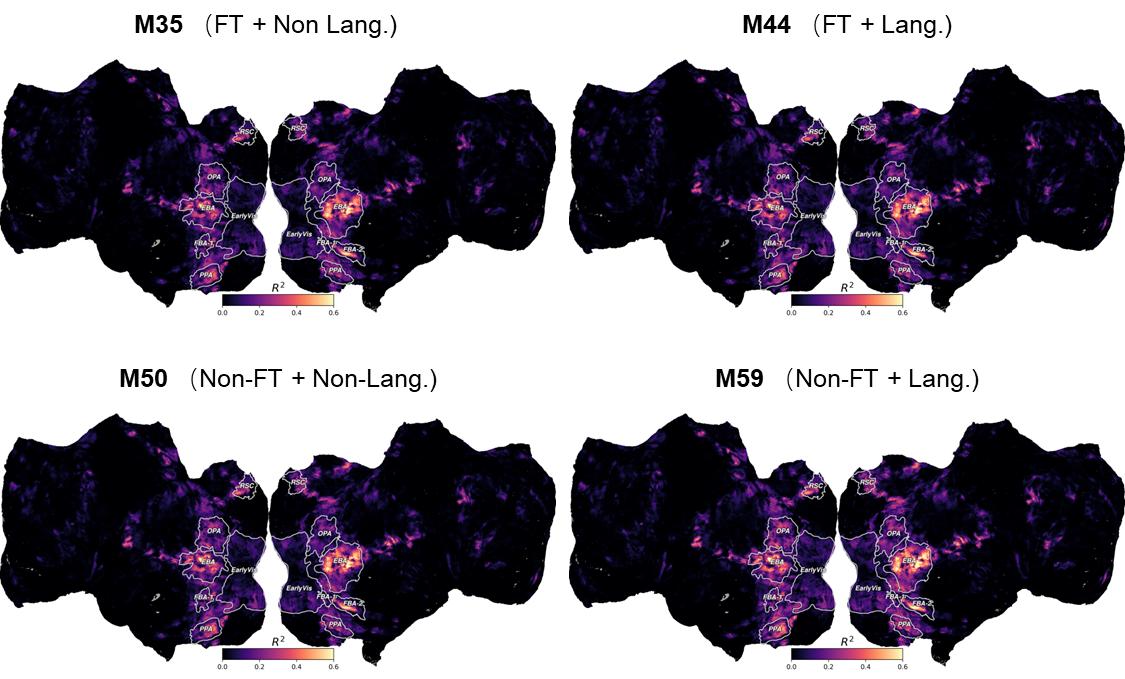}
  \caption{\textbf{Visualization of alignment performance across different models for Subject 1, focusing on the EBA region.} The four selected models vary in terms of whether they include language modality during pretraining and whether they are fine-tuned. Despite differences in overall alignment performance, a consistent pattern emerges where the highest alignment is observed in higher-level visual cortex regions, particularly in the EBA.}
  \label{fig:alignment_sameSubj_diffModels}
\end{figure}

Figure~\ref{fig:alignment_sameSubj_diffModels} illustrates the alignment performance across four different models for Subject~1. These models differ in two key aspects: the inclusion of language modality during pretraining and whether fine-tuning was applied. Despite variations in overall alignment performance, a consistent pattern emerges: higher alignment is consistently observed in higher-level visual cortex regions, particularly in EBA. This trend underscores the prominence of EBA in capturing higher-order visual semantics across model architectures, regardless of specific training configurations.

\begin{figure}[H]
  \centering
  \includegraphics[width=0.99\linewidth]{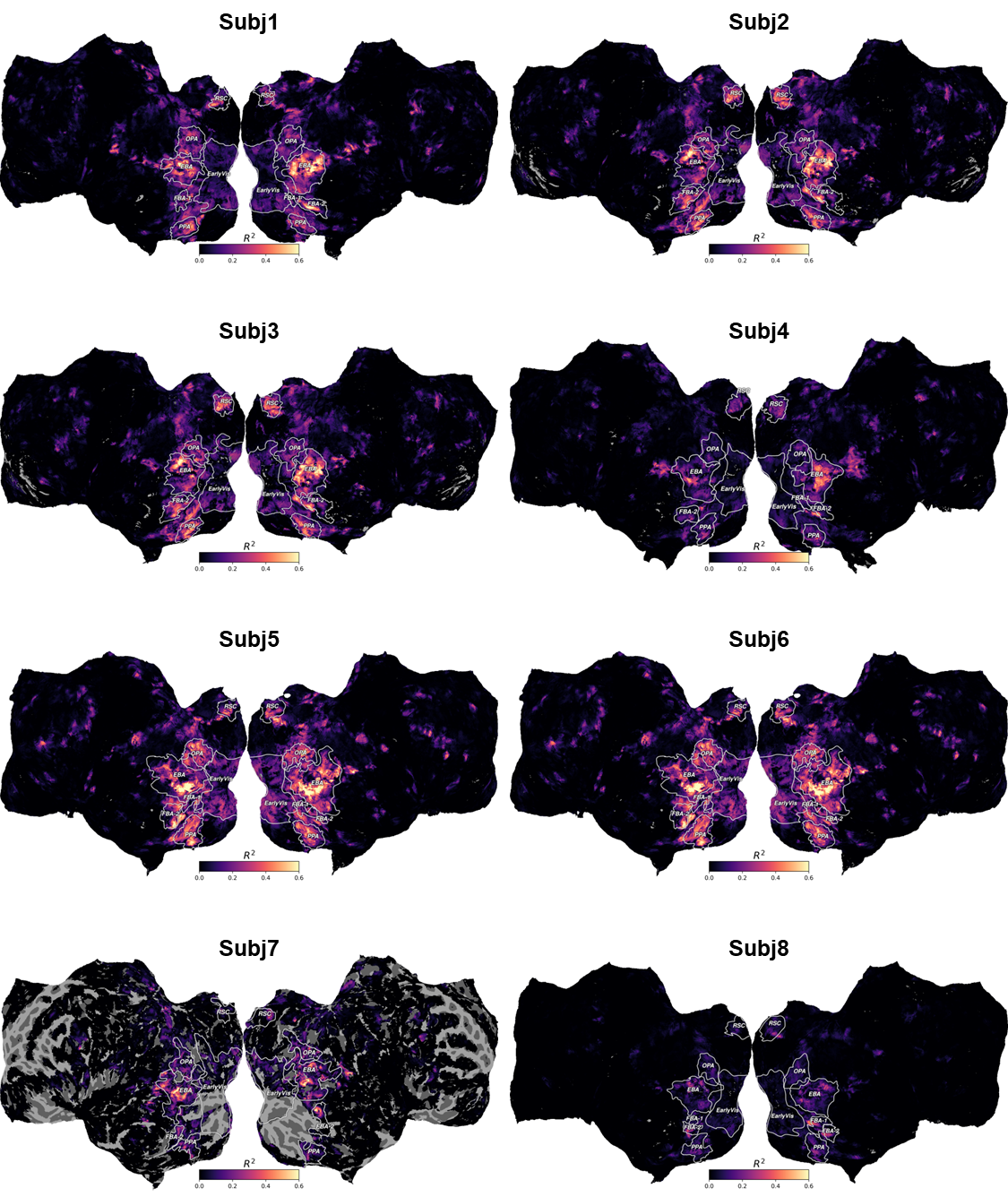}
  \caption{\textbf{Alignment performance across different subjects for the same model.} This figure illustrates the alignment performance of the same model across different subjects. There is a noticeable variation in alignment effectiveness, with Subjects 5 and 6 exhibiting better alignment, while Subjects 7 and 8 show relatively weaker alignment. This highlights substantial inter-subject variability in alignment performance.}
  \label{fig:alignment_sameModel_diffSubj}
\end{figure}

Figure \ref{fig:alignment_sameModel_diffSubj} illustrates the alignment performance of a selected model across the eight participants. Despite inter-subject variability, the overall spatial distribution of alignment remains relatively consistent across participants. Regions around the EBA consistently show higher alignment, indicating that model-derived representations align more effectively with neural activity in higher-order visual areas. This consistent spatial distribution suggests that the alignment patterns observed in the main analysis are not specific to a single participant but are generalizable across multiple subjects.

\section{Supplementary Analysis on Dimensionality}
\label{supple: Dimensionality}

\begin{figure}[H]
  \centering
  \includegraphics[width=0.95\linewidth]{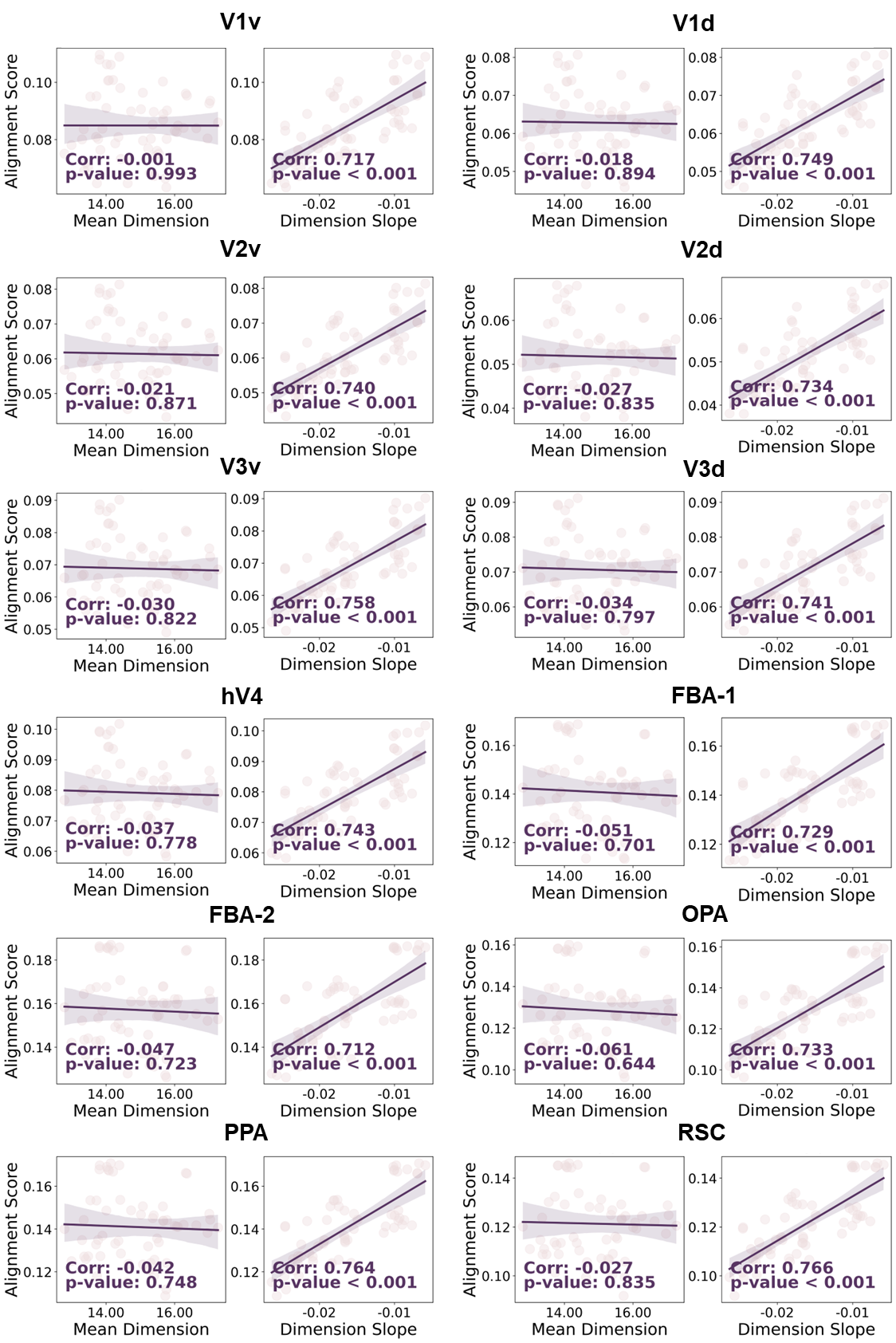}
  \caption{\textbf{Dimensional stability across different ROIs within the same participant (Subject 1)} The mean dimensionality and dimensional slope are calculated for each ROI to assess how structural consistency varies across brain regions. The results reveals that while the mean dimensionality shows limited predictive power in relation to alignment scores, the dimensional slope exhibits a strong negative correlation with alignment. 
}
  \label{fig:dimension_sameSubj_diffROIs}
\end{figure}

\begin{figure}[H]
  \centering
  \includegraphics[width=\linewidth]{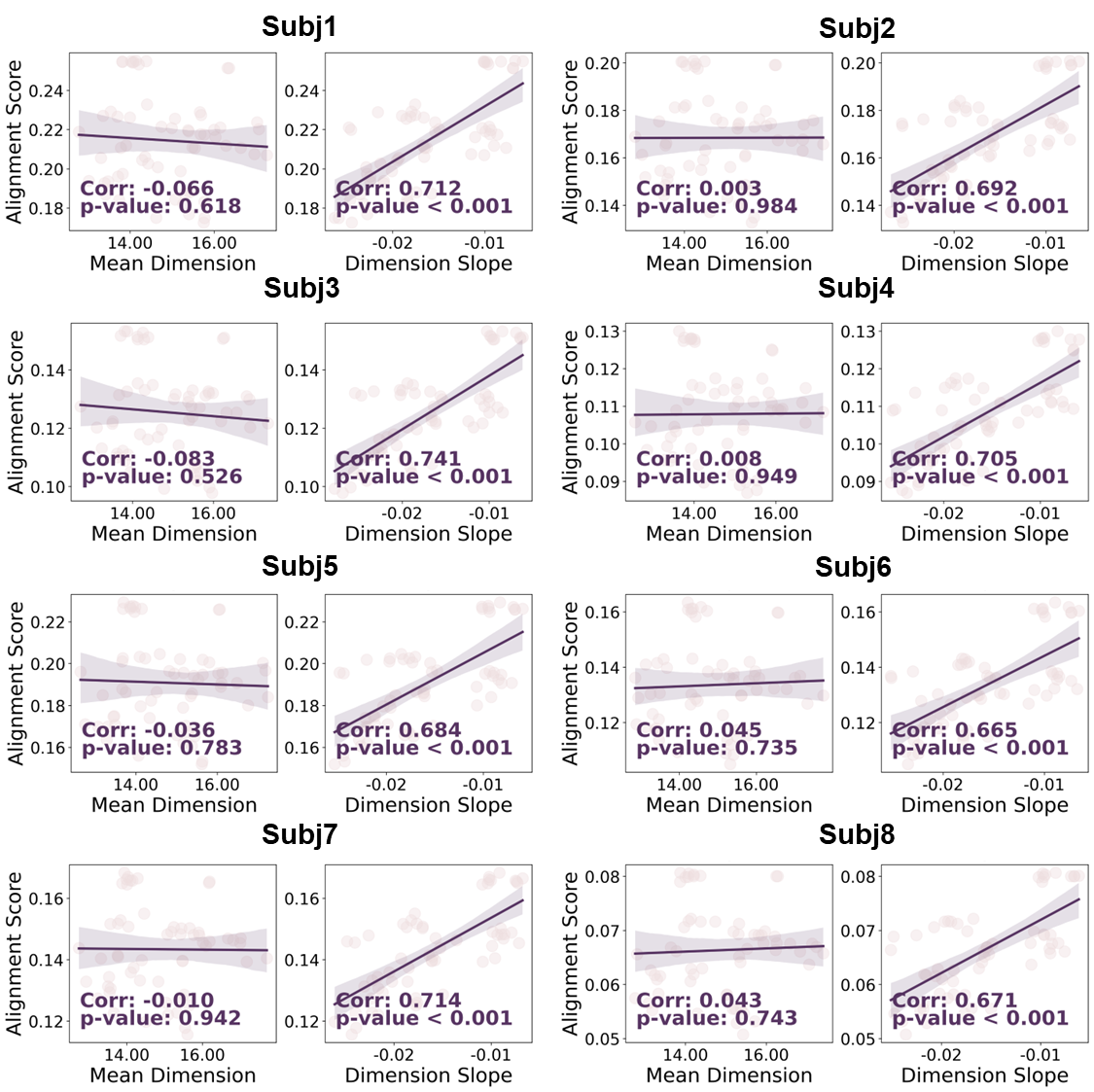}
  \caption{\textbf{Dimensional Analysis Across Subjects for the Same ROI (EBA)} This figure presents the dimensionality analysis for the embeddings within the EBA across multiple subjects. The mean dimensionality and dimensional slope are calculated based on the embeddings, remaining consistent across subjects, while alignment scores vary across individuals. The analysis highlights that while the mean dimensionality offers limited predictive power for alignment, the dimensional slope demonstrates a strong correlation with alignment scores, indicating that embeddings with more stable structural organization across scales tend to achieve stronger alignment with neural data. This consistency in embedding structure, despite inter-subject variability in alignment scores, further underscores the importance of scale-invariant structure in capturing brain-like representations.}
  \label{fig:dimension_sameROI_diffSubjs}
\end{figure}

We further extend the dimensionality analysis to encompass additional brain regions and participants, as illustrated in Figures \ref{fig:dimension_sameSubj_diffROIs} and \ref{fig:dimension_sameROI_diffSubjs}. These figures depict the relationship between mean dimensionality and alignment scores, as well as the correlation between dimensional slope and alignment scores across various regions and subjects.

Consistent with the main text, the mean dimension provides limited predictive power, while the dimensional slope shows a consistently strong correlation with alignment, reaching up to 0.766 in RSC . This further supports the conclusion that dimensional stability across scales is a key factor in capturing brain-like representations.

\section{Supplementary on Multi-scale Structural Self-Similarity Analysis}
\label{supple: Structural Self-Similarity}

\subsection{Multi-scale Structural Self-Similarity Analysis for More Subjects and Regions}

\begin{figure}[H]
  \centering
  \includegraphics[width=0.83\linewidth]{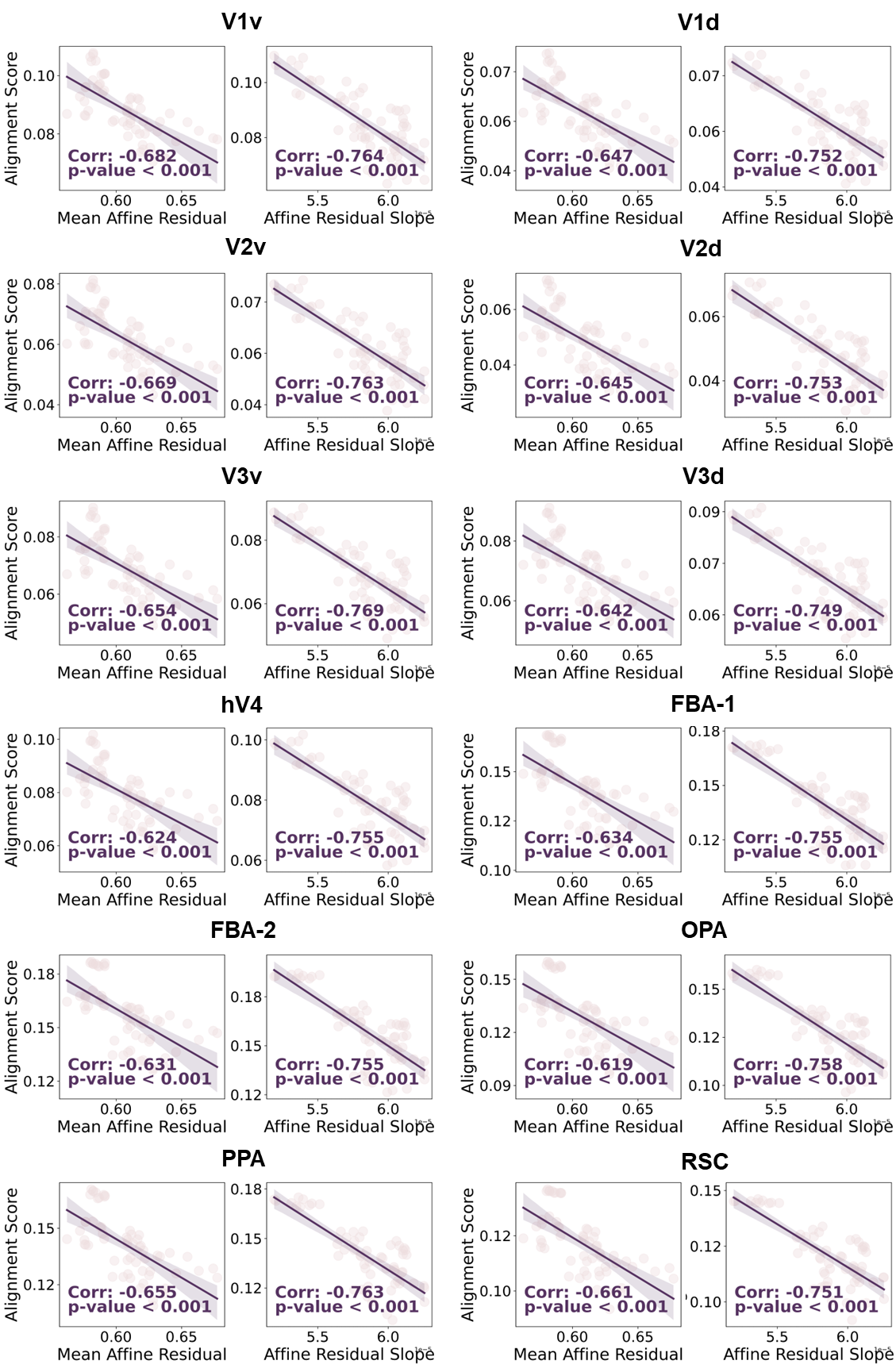}
  \caption{\textbf{Affine Residual Analysis Across Different ROIs.} This figure examines the relationship between multi-scale structural self-similarity and alignment performance across different brain regions for Subject 1. Results show that embeddings with lower mean affine residuals and flatter residual slopes exhibit stronger alignment with fMRI data, indicating that more consistent structural organization across scales is a key predictor of alignment performance.}
  \label{fig:affineResidual_sameSubj_diffROIs}
\end{figure}

\begin{figure}[H]
  \centering
  \includegraphics[width=\linewidth]{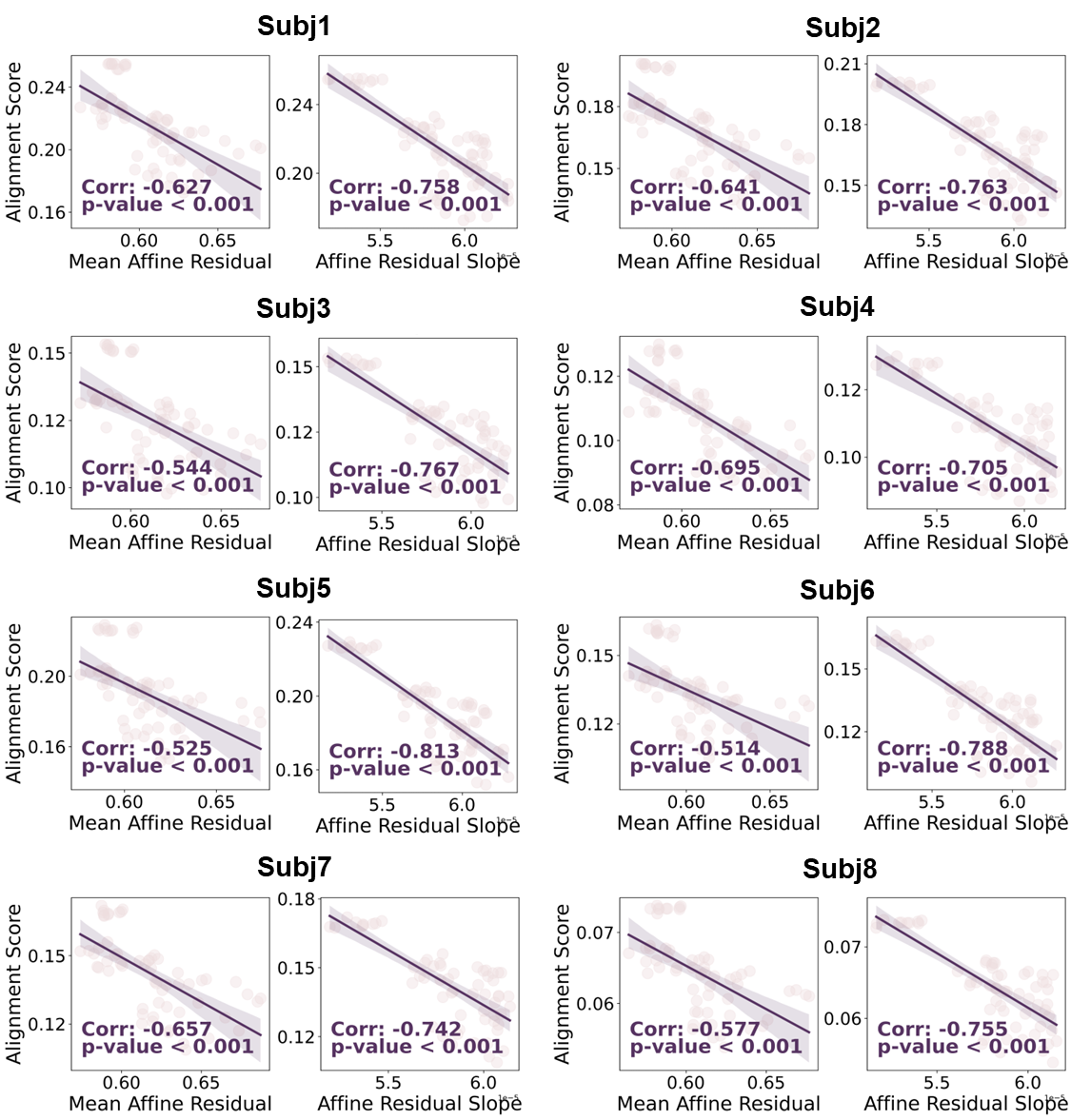}
  \caption{\textbf{Affine Residual Analysis for EBA Across Different Subjects.} This figure presents the affine residual analysis for the EBA region across multiple subjects. The Results reveal that embeddings with lower mean affine residuals and flatter residual slopes consistently exhibit stronger alignment across subjects. Despite inter-subject variability in alignment scores, the trend remains robust, indicating that maintaining structural consistency across scales is a key predictor of alignment regardless of individual differences.}
  \label{fig:affineResidual_sameROI_diffSubjs}
\end{figure}

In the main text, we focused on the Multi-scale Structural Self-Similarity Analysis within the EBA region. Here, we extend the analysis to additional brain regions and subjects to assess the generalizability of the observed patterns. By evaluating the affine residuals and their slopes across multiple regions and participants, we verify the consistency of the relationship between structural self-similarity and alignment performance. Figure \ref{fig:affineResidual_sameSubj_diffROIs} and Figure \ref{fig:affineResidual_sameROI_diffSubjs} confirm that embeddings with lower residuals and flatter slopes consistently align better with fMRI data across different cortical areas and subjects, reinforcing the importance of multi-scale structural self-similarity as a robust predictor of alignment.

\subsection{Multi-scale Structural Self-Similarity Analysis for More Scales}

To comprehensively assess structural consistency across different embedding scales, we extended the structural similarity analysis to include all pairwise scale combinations, focusing specifically on the EBA region. This expanded analysis provides a more granular view of how structural organization evolves across scales and how it relates to alignment performance.

\begin{figure}[H]
  \centering
  \includegraphics[width=1.0\linewidth]{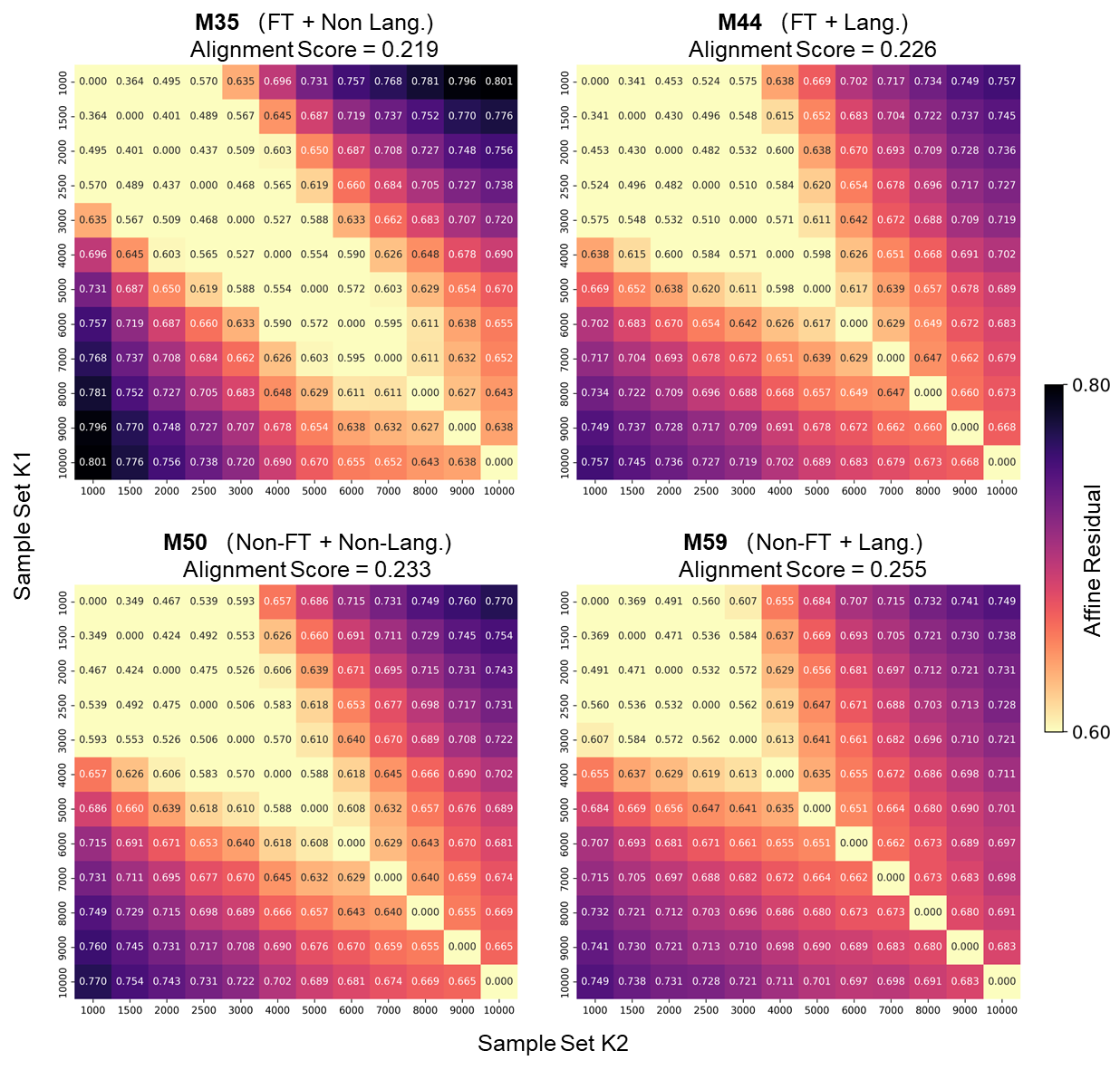}
  \caption{\textbf{Affine Residual Analysis for Selected Models with Varying Pretraining and Fine-Tuning Configurations.} This figure presents the affine residuals across scales for four selected models, strategically chosen to represent varying alignment performance and training configurations, including models with and without language modality during pretraining and with or without fine-tuning. Models pretrained with language modality and without fine-tuning generally exhibit lower affine residuals, indicating more stable structural organization across scales. In contrast, models without language modality or those fine-tuned for specific tasks show higher residuals, reflecting less structural consistency. Notably, the best-aligned model, which includes language modality and is not fine-tuned, maintains consistently low residuals across scales, underscoring the impact of pretraining configurations on structural coherence. The largest discrepancies in affine residuals are observed at broader scales, further emphasizing the importance of structural stability in achieving strong alignment.}
  \label{fig:affineResidual_4Examp}
\end{figure}

Figure \ref{fig:affineResidual_4Examp} presents the affine residuals across all scale combinations for four selected models. These four models were chosen to represent varying levels of alignment performance, specifically considering whether the models were fine-tuned and whether they included a language modality. The selected models span a range from the least aligned to the most aligned, providing a representative spectrum for examining structural variation.

From the results, we observe that models with lower alignment performance exhibit larger affine residuals across scales, indicating less structural consistency. As alignment performance improves, the maximum affine residual across scales decreases, suggesting that models with better alignment maintain more consistent structural organization across scales. Furthermore, the variance in affine residuals is noticeably lower in the best-aligned model, indicating a more stable and coherent structure throughout all scales.

\begin{figure}[H]
  \centering
  \includegraphics[width=1.0\linewidth]{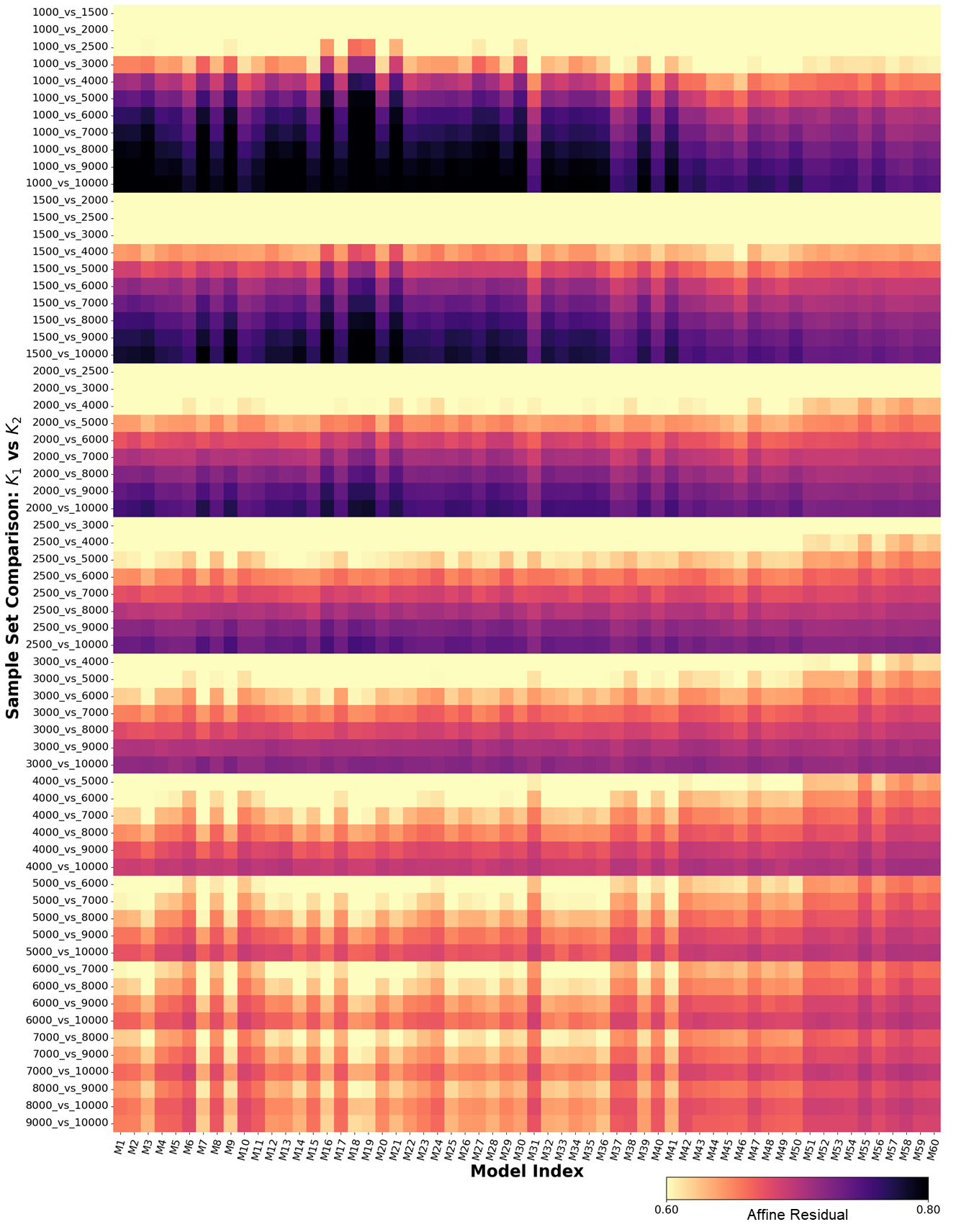}
  \caption{\textbf{Affine Residual Analysis Across All Models.} The figure presents the affine residuals for all 60 models, sorted by alignment performance from left to right. The analysis reveals a consistent trend where models with higher alignment exhibit lower affine residuals, indicating more stable structural organization across scales. Notably, the reduction in affine residuals is most pronounced at larger scales, suggesting that well-aligned models better preserve structural consistency across broader scales. This finding underscores the relationship between structural stability and alignment, with models that maintain more coherent multi-scale structures demonstrating stronger alignment performance .}
  \label{fig:affineResidual_60Examp}
\end{figure}

Figure \ref{fig:affineResidual_60Examp} extends this analysis to all 60 models, ordered by alignment performance from left to right. The plot reveals a consistent trend: as alignment improves, the structural similarity between the smallest and largest scales increases, evidenced by a reduction in affine residuals. Notably, the most pronounced differences are observed in the largest scales, where the residuals decrease more substantially for well-aligned models, indicating that these models better preserve structural consistency at broader scales.

\begin{figure}[H]
  \centering
  \includegraphics[width=1.0\linewidth]{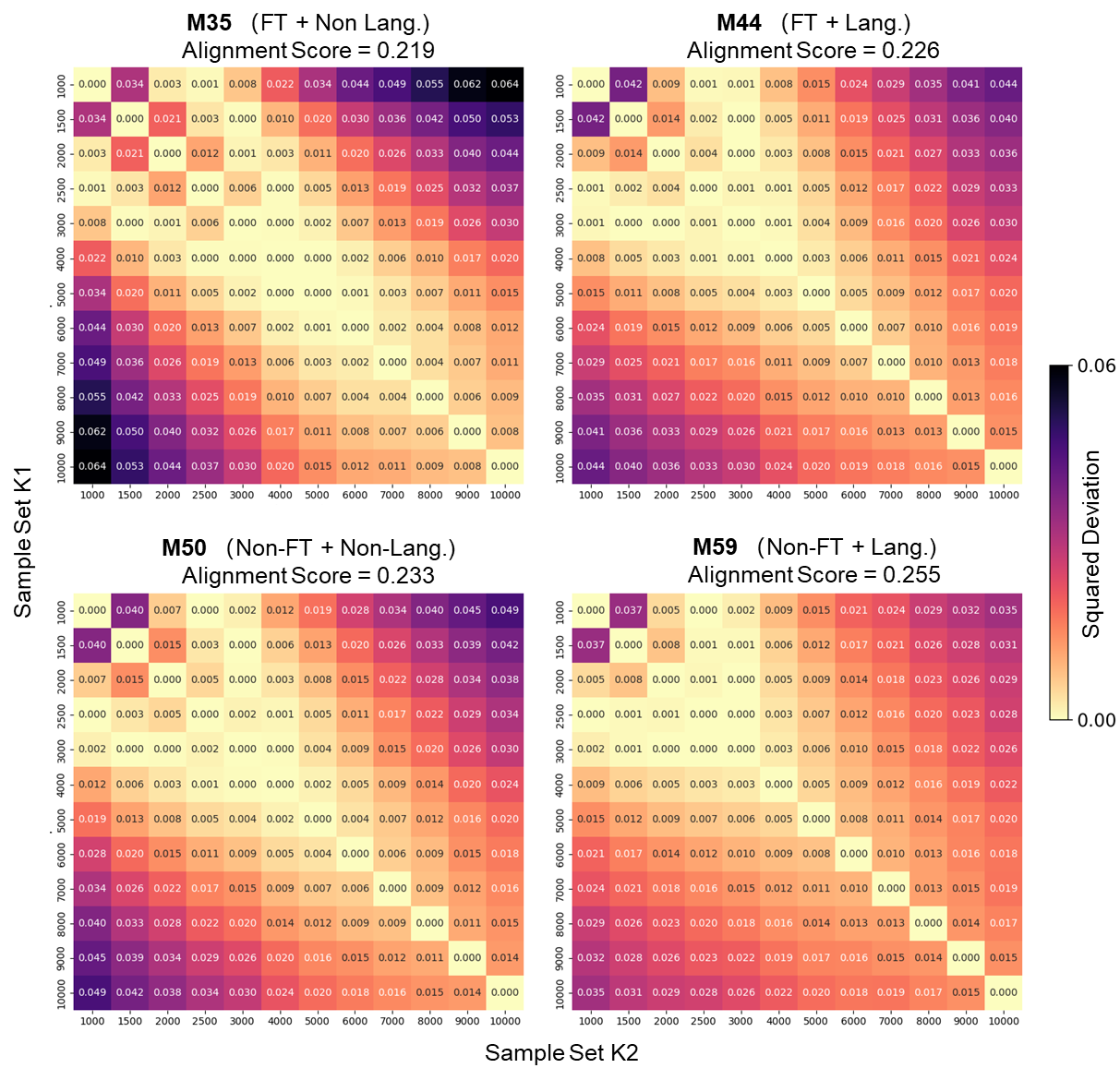}
  \caption{\textbf{Variance Analysis of Affine Residuals Across Selected Models.} This figure presents the variance of affine residuals across scales for the four selected models, chosen to cover a range of alignment performance and training configurations, including models with and without language modality and those with and without fine-tuning. The variance analysis provides deeper insights into the structural stability of each model across scales.
Models pretrained with language modality and without fine-tuning exhibit the lowest variance in affine residuals, indicating a more stable and coherent structural organization across scales. In contrast, models without language modality or those subjected to fine-tuning display significantly higher variance, suggesting greater structural fluctuations and less consistency across scales.
Notably, the variance differences are particularly pronounced at the largest scales, where the best-aligned model demonstrates minimal variance, further underscoring the impact of pretraining configurations on structural stability. This finding highlights that models with more stable multi-scale structures tend to maintain more consistent structural organization, aligning more effectively with brain-like representations.}
  \label{fig:SquaredDev_4Examp}
\end{figure}

To provide a more quantitative measure of structural variation, we calculated the variance of affine residuals across all scale combinations. Figure \ref{fig:SquaredDev_4Examp} illustrates this variance for the four selected models, showing a clear pattern where the best-aligned model exhibits the lowest variance in affine residuals, suggesting greater structural coherence across scales. The least-aligned model, in contrast, shows higher residual variance, indicating more structural deviation as scale changes.

\begin{figure}[H]
  \centering
  \includegraphics[width=1.0\linewidth]{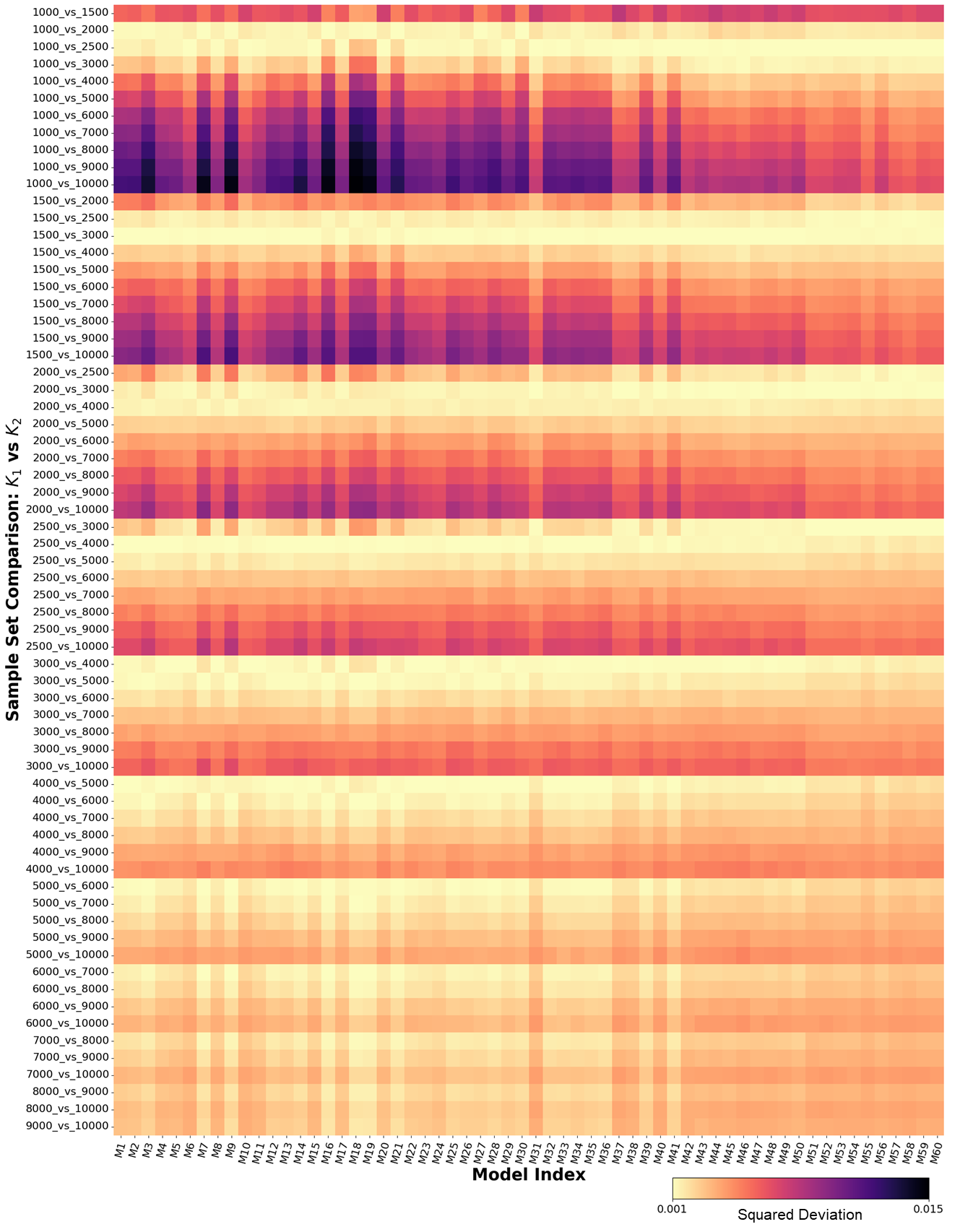}
  \caption{\textbf{Variance Analysis of Affine Residuals Across All Models, Ordered by Alignment Performance.} This figure presents the variance of affine residuals across all 60 models, systematically ordered by alignment performance from left (lowest alignment) to right (highest alignment). The analysis reveals a clear trend: as alignment performance improves, the variance of affine residuals consistently decreases. Models with lower alignment exhibit higher variance in affine residuals, indicating greater structural inconsistency across scales. In contrast, well-aligned models demonstrate significantly lower variance, reflecting more stable and coherent structural organization. Notably, the variance reduction is most pronounced at broader scales, suggesting that models with stronger alignment not only maintain more consistent structural organization but also better preserve multi-scale structural stability. This pattern reinforces the observation that structural consistency across scales is a key factor in achieving robust alignment with neural data.}
  \label{fig:SquaredDev_60Examp}
\end{figure}

Figure \ref{fig:SquaredDev_60Examp} presents the variance analysis for all 60 models, reinforcing the observed trend. The variance of affine residuals systematically decreases as alignment improves, providing a more robust indicator of structural stability across scales. This finding underscores that models with stronger alignment not only exhibit lower maximum residuals but also maintain more consistent structural organization across varying scales, reflecting a more stable embedding space.

\section{Supplementary Analysis on Persistent Homology}
\label{supple: Persistent Homology Results}

In this section, we extend the persistent homology analysis to examine structural scale distributions from two perspectives: across different brain regions and across different subjects.

\begin{figure}[H]
  \centering
  \includegraphics[width=0.8\linewidth]{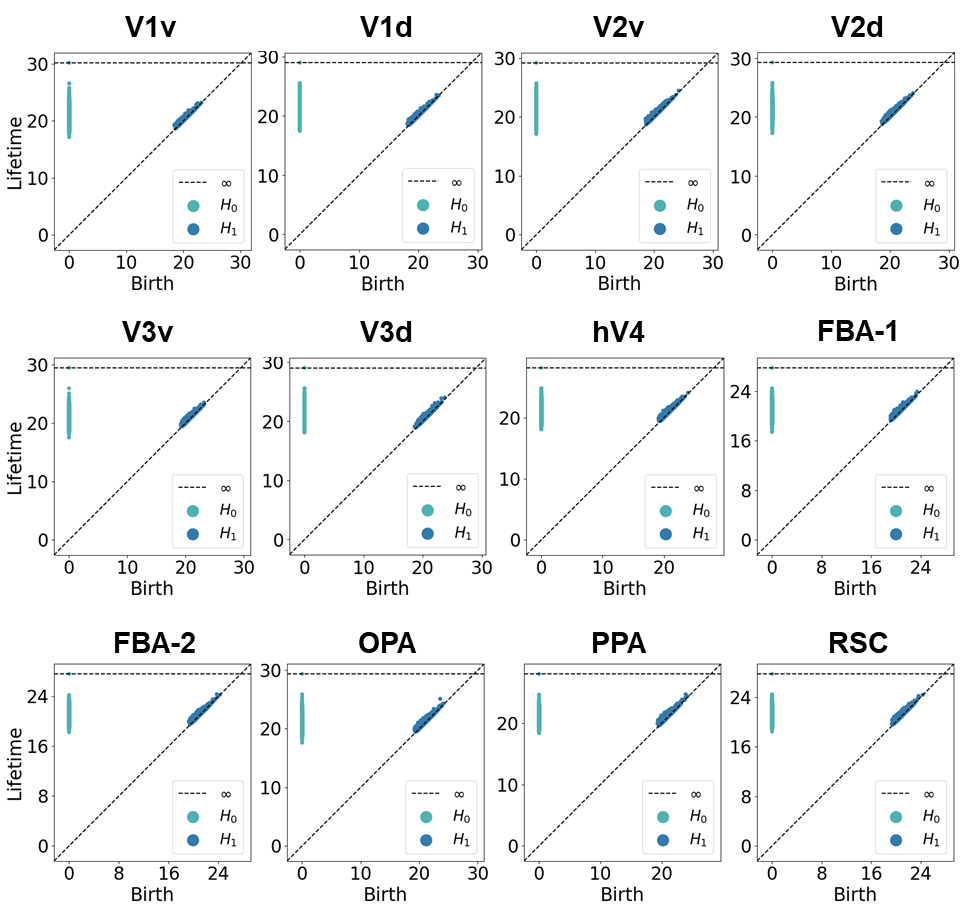}
  \caption{\textbf{Persistent Diagram Analysis Across Different ROIs.} This figure illustrates that the distribution of H1 features is highly consistent across regions, with most H1 features concentrated within a similar scale range (\textasciitilde20). In contrast, the death scales of H0 features vary more substantially, reflecting regional differences in the spatial extent of local clusters.}
  \label{fig:persistentDiagram_sameSubj_diffROIs}
\end{figure}

\paragraph{Analysis Across Brain Regions.}
We analyze the distribution of H1 features across multiple brain regions in Subject 1. Figure \ref{fig:persistentDiagram_sameSubj_diffROIs} shows that the H1 scale distributions are highly consistent across regions, with most H1 features concentrated around a similar scale range (\textasciitilde20), indicating stable structural patterns across cortical areas. In contrast, the death scales of H0 features vary more substantially, reflecting regional differences in the spatial extent of local clusters.  

\begin{figure}[H]
  \centering
  \includegraphics[width=0.8\linewidth]{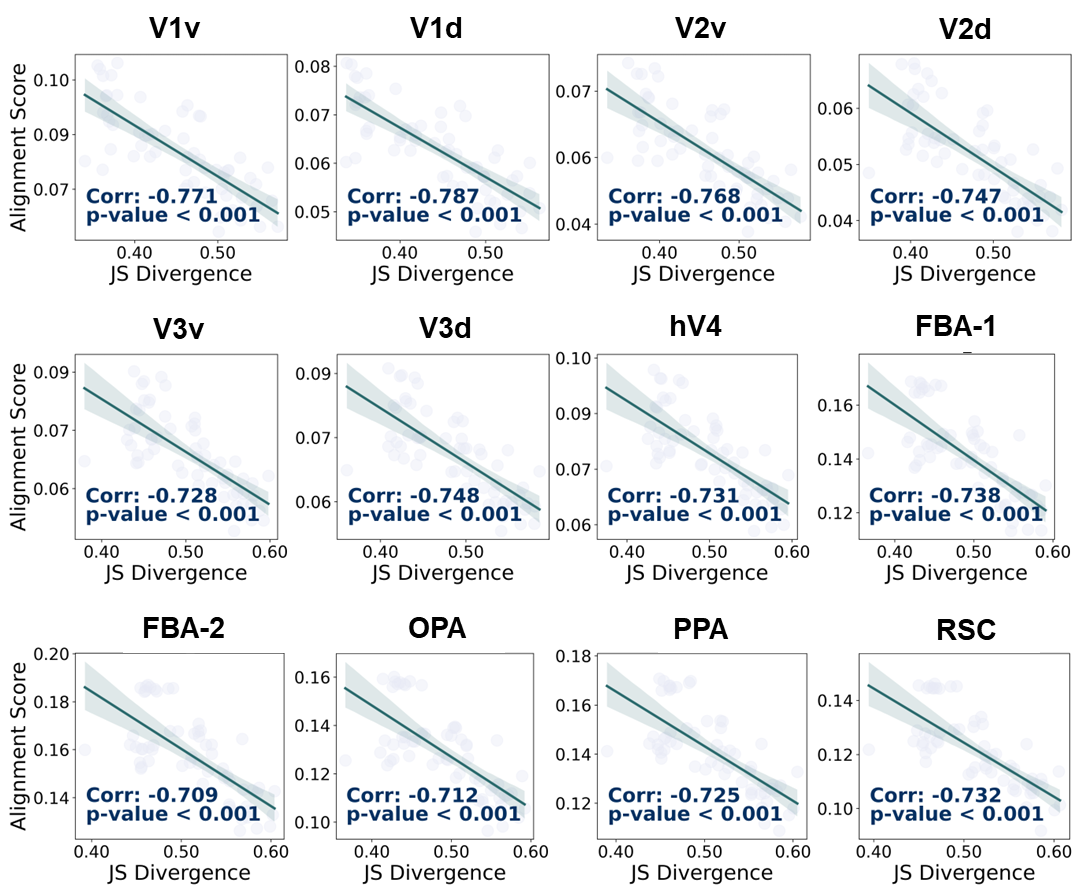}
  \caption{\textbf{Correlation Between JS Divergence and Alignment Score Across ROIs (Subject 1).} This figure illustrates the relationship between the JS divergence of H1 scale distributions and alignment scores across multiple brain regions in Subject 1. The analysis reveals a significant negative correlation, indicating that embeddings with H1 distributions more closely aligned with those of fMRI data (i.e., lower JS divergence) achieve better alignment performance.}
  \label{fig:PH_sameSubj_diffROIs}
\end{figure}

Additionally, we quantify the similarity between the embedding and fMRI H1 scale distributions using JS divergence and assess its correlation with alignment scores. Figure \ref{fig:PH_sameSubj_diffROIs} reveals a significant negative correlation, indicating that embeddings with H1 scale distributions more closely aligned with those of fMRI data (i.e., lower JS divergence) achieve better alignment performance. This further underscores the importance of structural scale consistency in predicting alignment effectiveness across different cortical regions.

\begin{figure}[H]
  \centering
  \includegraphics[width=0.8\linewidth]{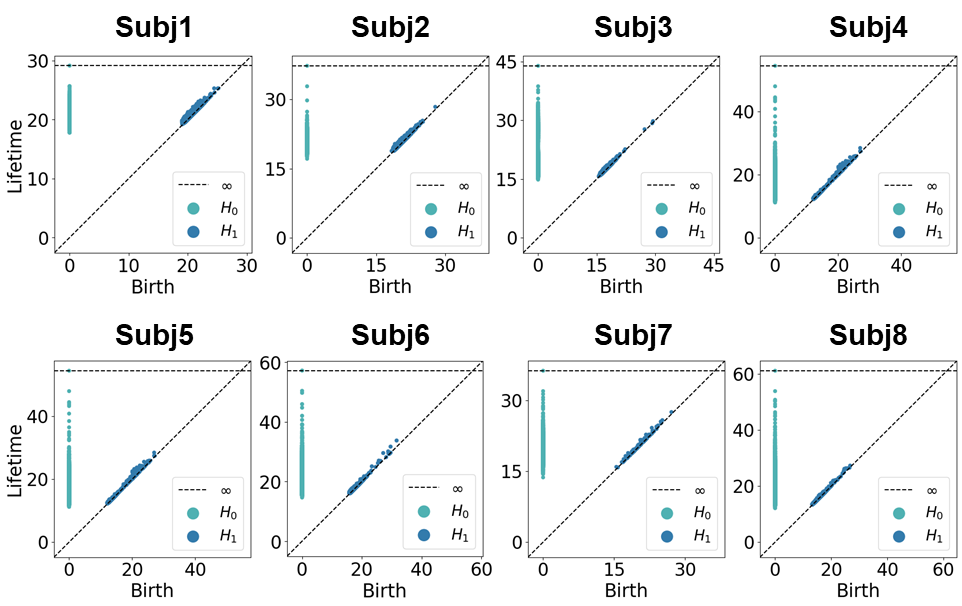}
  \caption{\textbf{Persistent Homology Analysis Across Subjects for the Same ROI (EBA).} This figure presents the distribution of topological features across scales in the EBA region for multiple subjects. The H1 scale distributions remain consistently concentrated around a similar scale (\textasciitilde20), indicating stable structural organization across subjects. In contrast, the death scales of H0 features show greater variability, reflecting individual differences in the spatial extent of local clusters.}
  \label{fig:persistentDiagram_sameROI_diffSubjs}
\end{figure}

\paragraph{Analysis Across Subjects.}
We assess the distribution of H1 features in the EBA region across multiple subjects. As shown in Figure \ref{fig:persistentDiagram_sameROI_diffSubjs}, the H1 scale distributions remain consistently concentrated around similar values (\textasciitilde20) across subjects, indicating stable structural organization despite individual variability. However, the death scales of H0 features exhibit greater variability, reflecting differences in the spatial extent of local clusters across subjects. 

\begin{figure}[H]
  \centering
  \includegraphics[width=0.8\linewidth]{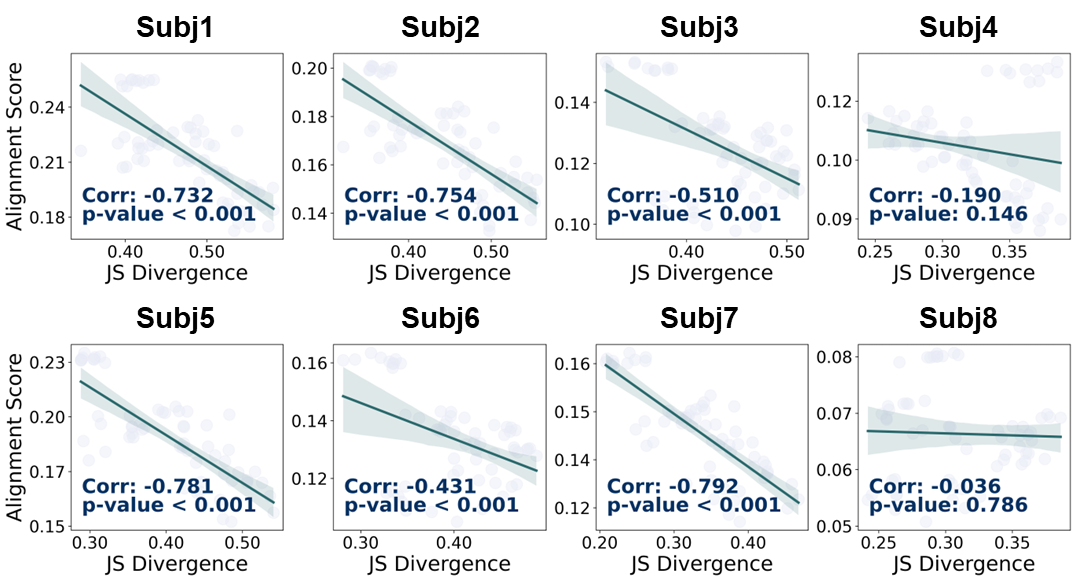}
  \caption{\textbf{Correlation Between JS Divergence and Alignment Score Across Subjects for EBA.} This figure examines the relationship between the JS divergence of H1 scale distributions in embeddings and alignment scores across multiple subjects for the EBA region. Overall, a significant negative correlation is observed, indicating that embeddings with H1 scale distributions more closely aligned with fMRI data (i.e., lower JS divergence) achieve better alignment performance. However, notable exceptions are observed for Subjects 4 and 8, where the JS divergence and alignment score show no significant correlation, suggesting potential individual differences in structural alignment patterns for these subjects.}
  \label{fig:PH_sameROI_diffSubjs}
\end{figure}

Figure \ref{fig:PH_sameROI_diffSubjs} further demonstrates a negative correlation between JS divergence and alignment scores, reinforcing that embeddings with H1 scale distributions closer to fMRI data achieve better alignment. This consistency across subjects highlights the robustness of H1 scale distribution alignment as a predictor of alignment performance, while the variability in H0 scales suggests subject-specific structural differences.

Overall, these analyses confirm that the H1 scale distributions are relatively consistent across both regions and subjects, while the H0 scales exhibit more variability, indicating structural differences in local cluster organization across cortical areas and individuals.

\section{Supplementary Analysis on Scale-Invariance in fMRI Data}
\label{supple: fMRI scale-variance}

In this section, we extend the scale-invariance analysis to the fMRI data from visual cortical regions, Subject 1, focusing on dimensional stability and structural self-similarity. Unlike AI embeddings, which are derived from pre-trained models, the fMRI data consist of voxel-wise responses to visual stimuli, providing a distinct perspective on neural manifold structures.

\begin{figure}[htbp]
  \centering
  \includegraphics[width=0.96\linewidth]{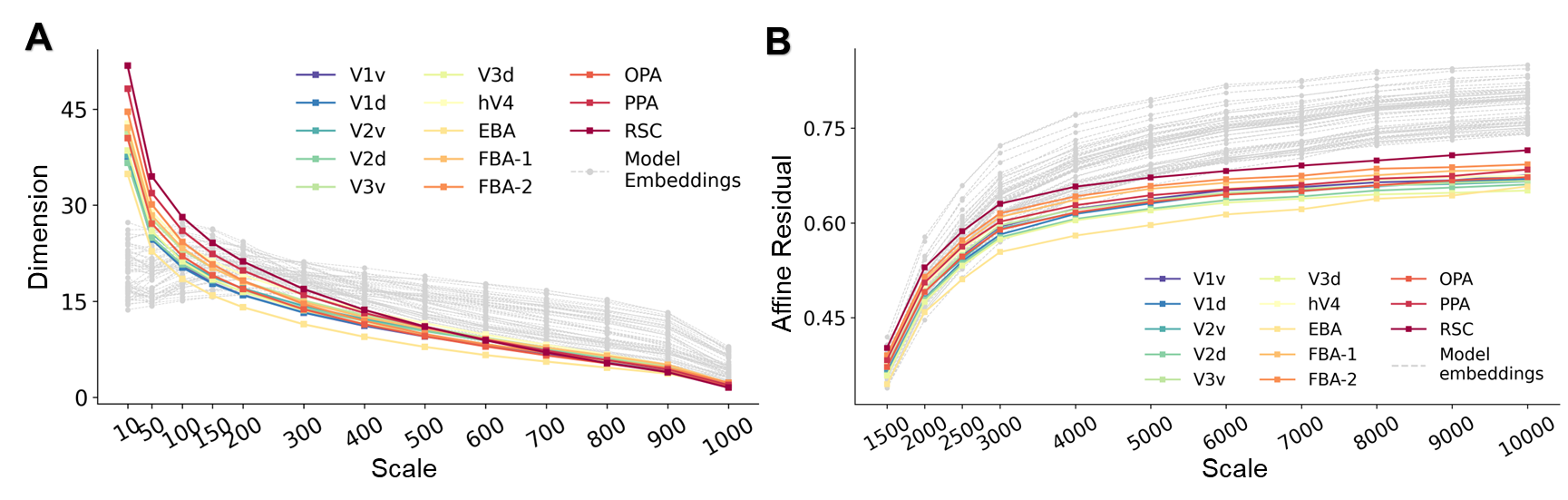}
  \caption{\textbf{Scale-Invariance Analysis of fMRI Data in Visual Cortex (Subject 1).} This figure shows the scale-invariance analysis of fMRI data in the visual cortex for Subject 1. At smaller scales, the dimensionality increases rapidly, indicating that local voxel-wise representations are less structured and more Euclidean-like. As the scale grows, the dimensionality stabilizes and remains lower than that of AI embeddings, suggesting that fMRI data maintain more compact representations at broader scales. In terms of structural self-similarity, fMRI data exhibit stronger consistency across scales compared to AI embeddings, particularly in visual processing regions. This highlights the stable structural organization in neural data across different scales.
}
  \label{fig:fMRIScaleInvariance}
\end{figure}

For each brain region, we extract voxel responses across all visual stimuli. Given that each participant views 10,000 images, we obtain a representation set of size $(10,000, \text{voxel\_num})$, where each row corresponds to the voxel-wise response to a single image stimulus. The voxel responses are flattened into a high-dimensional vector, effectively treating each region as a single high-dimensional representation space for subsequent analysis.

We analyze the dimensional stability of fMRI data by examining the dimensionality across scales (Figure \ref{fig:fMRIScaleInvariance}. At smaller scales, we observe a rapid increase in dimensionality across all brain regions, indicating that the manifold structures in fMRI data tend to degenerate into Euclidean-like spaces at finer scales. This trend suggests that local neighborhoods in fMRI data are less structured and more diffused, similar to Euclidean space.

However, as the scale increases, the dimensionality of fMRI data stabilizes and remains consistently lower than that of AI embeddings. This stability is particularly evident in mid-to-large scales, where fMRI data exhibit a more compact structural organization. This observation highlights that, despite local structural diffusion, fMRI manifolds maintain lower-dimensional representations over broader scales, contrasting with the higher and more variable dimensionality observed in AI embeddings.

To assess structural self-similarity, we evaluate the similarity between the smallest scale (finest resolution) and larger scales within each brain region. The analysis reveals that fMRI data exhibit stronger structural self-similarity compared to AI embeddings, with consistently lower affine residuals across scales. This pattern is observed across multiple brain regions, indicating that neural manifolds maintain more coherent structural organization even as the scale increases.

The enhanced structural self-similarity in fMRI data suggests that neural representations are more structurally aligned across scales, maintaining similar geometric configurations despite variations in scale. In contrast, AI embeddings exhibit greater structural variability, with more pronounced deviations in self-similarity as scale increases.

Our findings indicate that fMRI data exhibit distinct structural properties compared to AI embeddings:

\begin{itemize}[leftmargin=*]
    \item \textbf{Dimensional Stability:} While fMRI data show a rapid increase in dimensionality at small scales, the dimensionality stabilizes and remains consistently lower at mid-to-large scales, suggesting more compact manifold structures in neural data.
    \item \textbf{Structural Self-Similarity:} fMRI data maintain stronger structural self-similarity across scales, indicating more coherent structural organization, particularly at broader scales.
\end{itemize}

These results underscore the importance of scale-invariance in neural representations, with neural manifolds exhibiting more compact and consistent structures across scales compared to AI embeddings. This structural compactness and stability may serve as critical factors in achieving robust alignment between AI and neural data.

\newpage

\end{document}